\documentclass[onecolumn,12pt,journal,final]{IEEEtran}

\IEEEoverridecommandlockouts
\usepackage{amsfonts}
\usepackage[dvips]{graphicx}
\usepackage{times}
\usepackage{cite}
\usepackage{amsmath}
\usepackage{cases}
\usepackage{array}
\usepackage{dsfont}
\usepackage{amssymb}

\usepackage{stfloats}
\usepackage{slashbox}
\usepackage{graphicx}
\usepackage{footnote}
\usepackage{color}
\usepackage{booktabs}
\usepackage{array}
\usepackage{multirow}
\usepackage{bm}
\usepackage{empheq}
\usepackage[labelformat=simple]{subcaption}

\usepackage{amsthm}
\usepackage{algorithm}
\usepackage{algorithmic}
\usepackage{color}
\usepackage{bbm}

\theoremstyle{plain}
\newtheorem{theorem}{Theorem}

\newtheorem{lemma}{Lemma}
\newtheorem{proposition}{Proposition}
\theoremstyle{definition}
\newtheorem{example}{Example}
\newtheorem{definition}{Definition}
\newtheorem{remark}{Remark}
\usepackage{epstopdf}

\begin{document}
\abovedisplayskip=2pt 
\belowdisplayskip=2pt 

\title{Cache-Aided Interference Management in Partially Connected Linear Networks}
\author{\IEEEauthorblockN{Fan Xu, \IEEEmembership{Student Member,~IEEE}, Meixia Tao, \IEEEmembership{Fellow,~IEEE}, and Tiankai Zheng}\\
\thanks{The authors are with the Department of Electronic Engineering, Shanghai Jiao Tong University, Shanghai, China (Emails: xxiaof@sjtu.edu.cn, mxtao@sjtu.edu.cn, ht\underline{\;}zhengtiankai@sjtu.edu.cn).}
\thanks{This paper was presented in part in \cite{globecom}.}
}
\maketitle

\begin{abstract}
This paper studies caching in $(K\!+\!L\!-\!1)\!\times\!K$ partially connected wireless linear networks, where each of the $K$ receivers locally communicates with $L$ out of the $K\!+\!L\!-\!1$ transmitters, and caches are at all nodes. The goal is to design caching and delivery schemes to reduce the transmission latency, by using normalized delivery time (NDT) as the performance metric. For small transmitter cache size (any $L$ transmitters can collectively store the database just once), we propose a cyclic caching strategy so that each of every $L$ consecutive transmitters caches a distinct part of each file; the delivery strategy exploits coded multicasting and interference alignment by introducing virtual receivers. The obtained NDT is within a multiplicative gap of 2 to the optimum in the entire cache size region, and optimal in certain region. For large transmitter cache size (any $L$ transmitters can collectively store the database for multiple copies), we propose a modified caching strategy so that every bit is repeatedly cached at consecutive transmitters; the delivery strategy exploits self-interference cancellation and interference neutralization. By combining these schemes, the NDT is optimal in a larger region. We also extend our results to linear networks with heterogeneous receiver connectivity and partially connected circular networks.
\end{abstract}

\begin{IEEEkeywords}
Wireless cache network, partially connected network, coded caching, and interference management.
\end{IEEEkeywords}


\section{Introduction}
Caching is a novel solution to leverage the increasingly rich storage resource for enhancing communication efficiency in wireless networks. Its main idea is to prefetch popular video contents at the network edge during off-peak times so as to smoothen the peak-hour traffic congestion and reduce user access latency \cite{liu2014content,femtocaching_and_d2d}. The promise of caching mainly owes to the invention of coded caching in an information-theoretic framework \cite{fundamentallimits}. It is revealed in \cite{fundamentallimits} that, in addition to the local receiver caching gain, coded caching allows a global caching gain that scales with the accumulated cache size in the network. This global caching gain is exploited through proper file partition for cache placement and then coded multicast transmission simultaneously useful for multiple user requests.

While coded caching is originally proposed for ideal error-free shared links with receiver caches, its extension to wireless interference networks with fading has been studied in \cite{upperbound,mine,niesen,bothcache,gunduz,cao}. Specifically, the authors in \cite{upperbound} first studied a $3\times3$ interference channel with caches only at the transmitter side. They showed that transmitter caches can turn the interference channel into more favorable channels including X channel and broadcast channel by proper file splitting and placement strategies, thereby obtaining  higher degrees of freedom (DoF) of the channel. Follow-up works have extended \cite{upperbound} to the network with caches at both transmitter and receiver sides. The authors in \cite{mine} characterized the upper and lower bounds of the normalized delivery time (NDT) of a general interference network with arbitrary number of transmitters, arbitrary number of receivers, and at any feasible cache size region (feasible in the sense of collectively storing the entire content library). This work reveals that joint transmit/receive coded caching can turn the interference network into a new class of channels,  namely, cooperative X-multicast channels.  The achievable NDT of the network can thus be minimized by optimizing the combination of interference alignment gain \cite{Xchannel}, interference neutralization gain \cite{zeroforcing}, coded multicasting gain \cite{fundamentallimits}, as well as receiver local caching gain at different cache sizes. In \cite{niesen}, by separating the physical layer and the network layer, the authors obtained an order-optimal approximation of system DoF through interference alignment. The authors in \cite{bothcache} and \cite{gunduz} analyzed the standard sum DoF, by a one-shot linear interference neutralization scheme and by a combination of interference alignment, zero-forcing and interference cancellation techniques, respectively. The authors in \cite{cao} extended the study in \cite{mine} to the multi-input multi-output (MIMO) interference network where each node is equipped with multiple antennas. It is found that the MIMO multiplexing gain and the caching gain are cumulative. Coded caching has also been extended in cloud-aided wireless interference networks, e.g., \cite{simeone,yuan,girgis2017decentralized,TaoInvited}, often referred to as fog radio access networks (Fog-RANs), to reduce the overall content delivery latency.

Note that all the above information-theoretic studies on cache-aided interference networks in \cite{upperbound,mine,cao,niesen,bothcache,gunduz} assumed a fully connected network topology, where all the transmitters can communicate with all the receivers over independent and identically distributed (i.i.d.) fading channels. In practical wireless networks, considering the path loss resulting from radio propagation and signal attenuation due to blocking objects, some links are inevitably weaker than others. This scenario is typically modeled as \emph{partially connected network} in the literature, where each receiver can only communicate with a subset of transmitters. With partial connectivity, each receiver incurs less interference but also can access smaller cumulative transmitter cache size. It is thus of both theoretical and practical importance to investigate the impact of caching in partially connected interference networks.
The works \cite{jimingyue,tangli,yener,wankai} studied caching in combination networks, a class of networks defined in \cite{combinationnetwork}  for network coding, where a single server is connected to $\binom{K}{n}$ receivers through a layer of $K$ relay nodes, such that each receiver is connected to a unique subset of $n$ relay nodes. Such networks, though exhibiting partial connectivity, are far from practical wireless interference networks as each connection link is dedicated and error-free. The authors in \cite{SmallCellCaching,KakarCacheHetNet} considered a cloud-aided $2\times2$ network with edge caching, where the connectivity is modeled as \emph{Z-connectivity}. Specifically, the authors in \cite{SmallCellCaching,KakarCacheHetNet} used an on-off interference channel and a linear deterministic model, respectively,  to characterize the channel between the edges and the receivers. Both works obtained the optimal delivery time per bit of the considered network models by exploiting self-interference cancellation and interference neutralization. These schemes however cannot be extended to a general partially connected network with arbitrary number of nodes.  The authors in \cite{topologicalcaching} considered a given but arbitrarily partially connected network with transmitter caches where the transmitters have no access to channel state information (CSI) beyond the network connectivity. They formulated a joint file placement and delivery optimization problem without information-theoretical optimality analysis. The most related work to our considered partially connected wireless interference networks is \cite{Wynercaching}. It considered the Wyner's network model where each cache-aided receiver is connected to two or three nearby transmitters circularly and each transmitter has no cache but can only access the files desired by its connected receivers. By using self-interference cancellation facilitated by receiver caches and local cooperation across transmitters connecting common receivers, the authors obtained an achievable rate-memory tradeoff which is optimal in certain cache size region.

In this paper, we study cache-aided partially connected \emph{linear} interference networks, where there are $K$ linearly aligned receivers, $K+L-1$ linearly aligned transmitters, and each receiver is locally connected to a subset of $L$ consecutive transmitters. This network model can be viewed as an extension of the Wyner's model in \cite{wyner,Wynercaching}. It can well model many practical communication systems such as highway roadside communications and railway wayside communications. Our goal is to find the minimum achievable NDT of this network with partial connectivity and cache sizes as key network parameters. We also aim to shed light on cache-aided interference management techniques in more practical wireless interference networks. While part of this work is presented in \cite{globecom}, the main contribution of this work is given as follows.
\begin{itemize}
  \item First, we propose a \emph{basic} caching and delivery scheme for small transmitter cache size, where the accumulated cache size at every set of $L$ transmitters is just enough to collectively store the entire database (i.e. $L \mu_T=1$ with $\mu_T$ being the fraction of the database that each transmitter can cache locally). A cyclic caching (CC) strategy is proposed at both the transmitter and receiver sides based on pure file partition that takes the partial linear connectivity into account. Each of every $L$ consecutive transmitters caches a distinct part of each file, and the entire database is recoverable by combining the cached content of every $L$ consecutive transmitters. In the delivery phase, by introducing virtual receivers, we exploit both coded multicasting gain and interference alignment gain. By using memory sharing between the considered case, i.e., $\mu_T=\frac{1}{L}$, and the trivial case where each receiver can cache the entire database, we obtain an achievable NDT in the entire feasible cache size region $L\mu_T+\mu_R\ge1$, with $\mu_R$ being the fraction of the database that each receiver can cache locally. This NDT is proved to be within a multiplicative gap of 2 to the information-theoretic optimum in the entire cache size region, and exactly optimal when $\mu_R\ge\max\{1-\mu_T,\frac{L-1}{L}\}$.
  \item Second, we propose an \emph{enhanced scheme} for large transmitter cache size where the accumulated cache at every set of $L$ transmitters can collectively store the entire database for $p\ge2$ times (i.e., $L\mu_T = p$). The proposed caching strategy of this enhanced scheme builds upon the previous CC strategy, referred to as modified CC strategy, so that every $p$ consecutive transmitters cache a common part of each file. In the delivery phase,  by utilizing the overlapped transmitter cache content, we achieve the combined gain of self-interference cancellation and interference neutralization. This enhanced scheme at large transmitter cache size in conjunction with the basic scheme at small transmitter cache size  allows us to obtain a lower achievable NDT which is exactly optimal when $\mu_T+\mu_R\ge1$.
  \item Finally, we extend our results to other partially connected networks. Our results are directly applicable to circular networks when $L$ is a divisor of $K$. We also extend our basic scheme to linear networks with heterogeneous receiver connectivity, and obtain an achievable NDT within a bounded multiplicative gap to the optimum.
\end{itemize}

The main novelty of our proposed achievable schemes lies on the appropriate combination of several interference management techniques, namely, interference alignment, interference neutralization, self-interference cancellation, as well as coded multicasting, facilitated by proper caching strategies. These techniques are previously well investigated in fully connected interference networks as summarized in \cite{TaoInvited}, but are not clearly understood in partially connected networks. The related work \cite{Wynercaching} only achieved self-interference cancellation gained by receiver caches and only allowed local cooperation across transmitters connecting common receivers due to the lack of transmitter cache.

The analytical findings in our work reveal several insights on caching in partially connected linear networks:
\begin{itemize}
  \item Transmitter cooperation via interference neutralization (facilitated by our proposed modified CC strategy) can achieve the optimal NDT as long as the one transmitter and one receiver together can collectively store the entire database, i.e. $\mu_T+\mu_R\ge1$, though it cannot reduce NDT by more that a constant factor of 2.
  \item Pure file splitting is globally optimal when $\mu_T+\mu_R\ge1$, and inter-file or intra-file coding can only improve NDT within a constant factor of 2 when $\mu_T+\mu_R<1$.
  \item Numerical results show that when   the receiver connectivity $L$ increases, our achievable NDT slightly decreases if $\mu_R<1-\mu_T$, and remains fixed if $\mu_R=0$ or $\mu_R\ge1-\mu_T$. This implies that, when the connectivity increases, the positive impact of having more transmitter cooperation or coordination slightly prevails the negative impact of more interferences received at each receiver if $0<\mu_R<1-\mu_T$, and offsets the negative impact otherwise.
\end{itemize}

After the prior conference publication of this work \cite{globecom}, the works \cite{gunduz_random,gunduz_partial,AliCellular} studied the similar problem of cache-aided partially connected wireless networks. Specifically, the authors in \cite{gunduz_random} studied caching in the network with random connectivity. A caching strategy using maximum distance separable (MDS) coding was proposed, and both successive and parallel transmissions were adopted. We shall compare our results with \cite{gunduz_random} in Section \ref{sec comparison}. The authors in \cite{gunduz_partial} studied receiver caching in a cloud-aided partially connected interference network where the partial connectivity is similar to that in combination networks but with channel fading. They obtained the achievable NDT for $L=2$ and arbitrary $\mu_R$, and for arbitrary $L$ when $\mu_R$ is sufficiently large. The authors in \cite{AliCellular} studied caching in a hexagonal cellular network where each user is only able to receive signals from its three neighboring base stations. By using a randomized caching strategy and a one-shot linear delivery strategy \cite{bothcache}, they obtained an achievable per-cell DoF which is within a constant gap to the optimum.

The rest of this paper is organized as follows. Section \ref{section model} introduces our system model and performance metric. Section \ref{section alignment} presents the basic scheme and its achievable NDT. Section \ref{section neutralization} presents the enhanced scheme and the achievable NDT by combining the two schemes. Section \ref{sec circular} extends our results to the circular networks and the linear networks with heterogeneous receiver connectivity. Finally, Section \ref{section conclusion} concludes this paper.

Notations: $[K]$ denotes the set $\{1,2,\ldots,K\}$, and $[J:K]$ denotes the set $\{J,J+1,\ldots,K-1,K\}$ for $J<K$. $\mathcal{CN}(0,1)$ denotes the Gaussian distribution with zero mean and unit variance. $\mathbf{P}\prec\mathbf{Q}$ denotes that the linear space spanned by the column vectors of matrix $\mathbf{P}$ is a subspace of the one spanned by the column vectors of matrix $\mathbf{Q}$.

\section{System Model}\label{section model}
\subsection{Network Model}
\begin{figure}[tbp]
\begin{minipage}[t]{1\linewidth}
\centering
\includegraphics[scale=0.15]{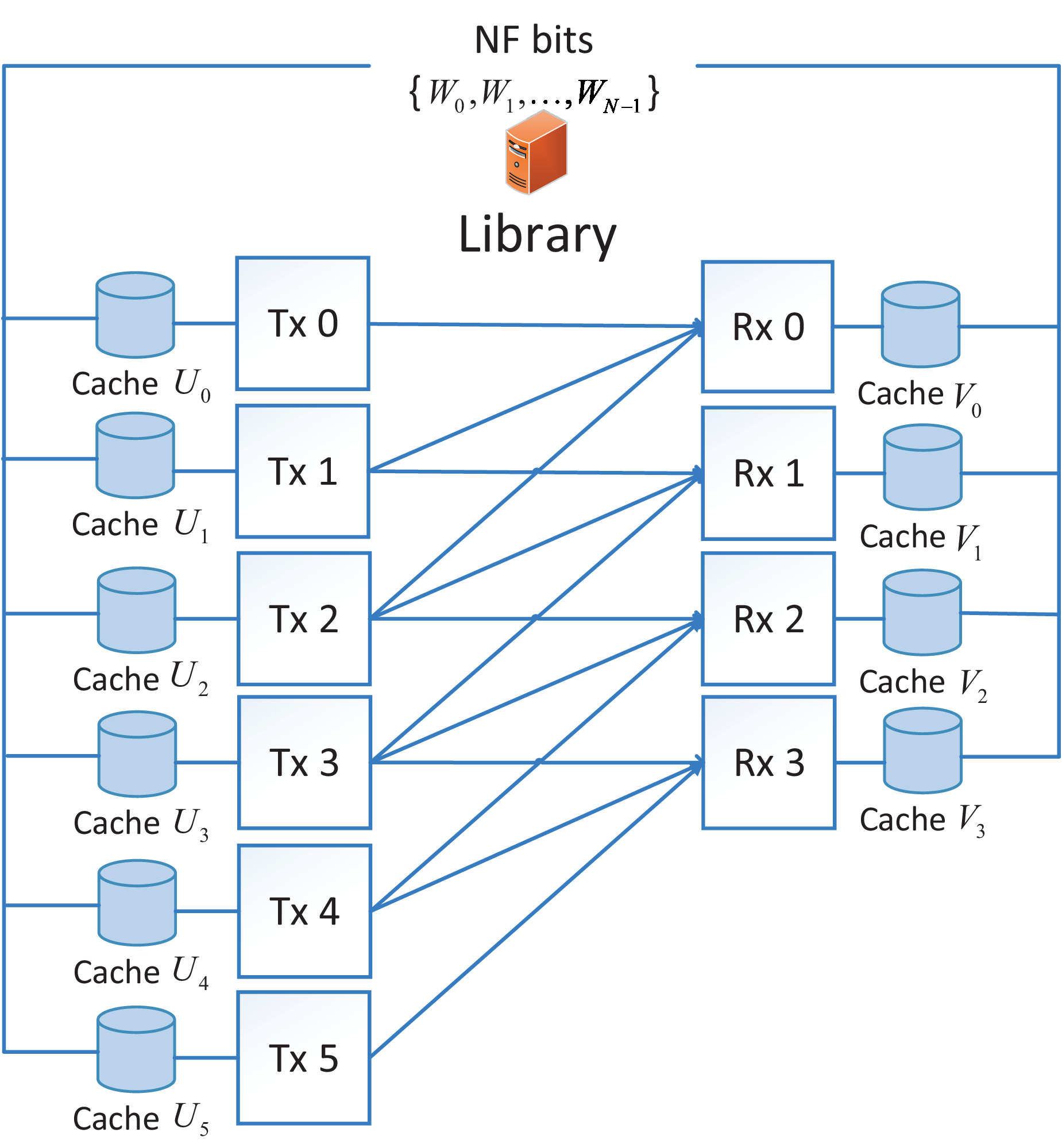}
\subcaption{}\label{Fig linear model}
\end{minipage}
\begin{minipage}[t]{1\linewidth}
\centering
\includegraphics[scale=0.15]{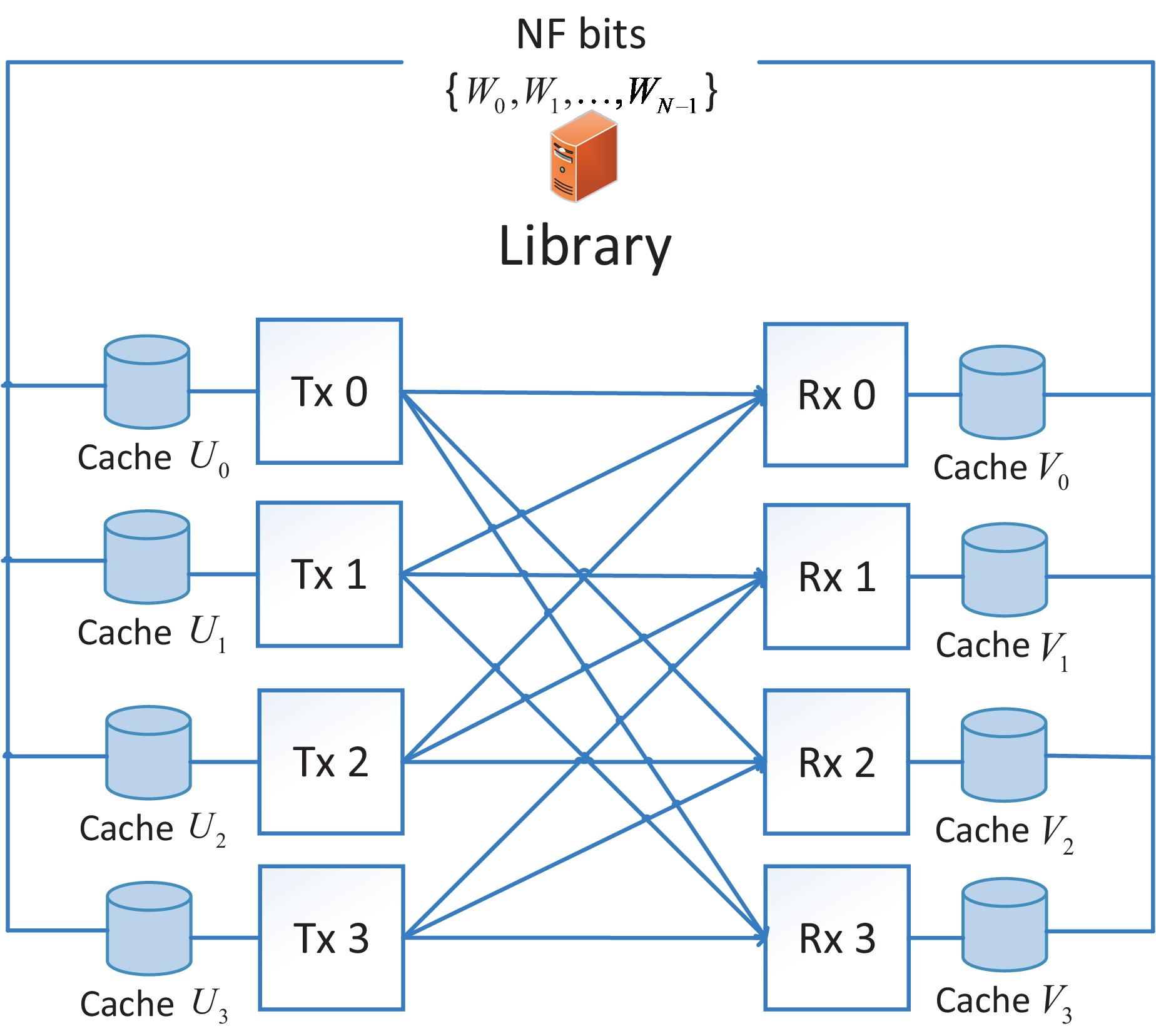}
\subcaption{}\label{Fig circular model}
\end{minipage}
\caption{Partially connected linear interference networks with receiver connectivity $L=3$: (a) $6\times 4$ linear network, (b) $4\times 4$ circular network.}\label{Fig partial model}
\end{figure}

We consider a $(K+L-1)\times K$ partially connected linear interference network, where there are $K+L-1$ linearly aligned transmitters, indexed by $\{0,1,\ldots,K+L-2\}$, $K$ linearly aligned receivers, indexed by $\{0,1,\ldots,K-1\}$, and each receiver $i$ is connected to $L$ consecutive transmitters $\{i,i+1,\ldots, i+L-1\}$, with $L\le K$. Here, $L$ is referred to as \emph{receiver connectivity}. Fig.~\ref{Fig linear model} shows an example of the linear network with  $K\!=\!4$ and $L\!=\!3$.  Let $\mathcal{T}_i\triangleq[i:i+L-1]$, for $i\!\in\![0\!:\!K\!-\!1]$, denote the set of transmitters connected to receiver $i$, and let $\mathcal{R}_j\!\triangleq\![j\!:\!j\!-\!L\!+\!1]\cap[0\!:\!K\!-\!1]$, for $j\!\in\![0\!:\!K\!+\!L\!-\!2]$, denote the set of receivers connected to transmitter $j$. Note that while $|\mathcal{T}_i|=L$, $|\mathcal{R}_j| \le L$. Each node is equipped with a cache memory of finite size, and has single antenna. The communication at each time slot $t$ over this network is modeled by
\begin{align}
Y_i(t)=\sum_{j\in\mathcal{T}_i}h_{i,j}(t)X_j(t)+Z_i(t),\notag
\end{align}
where $Y_i(t)\in \mathbb{C}$ denotes the received signal at receiver $i\in[0:K-1]$, $X_j(t)\in \mathbb{C}$ denotes the transmitted signal at transmitter $j$, $h_{i,j}(t)\in \mathbb{C}$ denotes the channel coefficient from transmitter $j$ to receiver $i$ which is i.i.d. for different $i$, $j$, and $t$, following some continuous distribution, and $Z_i(t)$ denotes the noise at receiver $i$ distributed as $\mathcal{CN}(0,1)$. At each time slot $t$, all the transmitters estimate their local CSI and aggregate them at a central controller. The central controller uses the current global CSI to design the precoding coefficient of each transmitted symbol for each transmitter at this time slot. When each receiver receives all its signals at all time slots, the central controller then computes the receive combining matrix for each receiver based on all the precoding coefficients and all the channel realizations during the transmission.

This network model is an extension of the $K\times K$ partially connected interference network in \cite{VVVpartialnetwork}, where there are $L-1$ more transmitters to maintain the constant connectivity of $L$ for all receivers. A constant receiver connectivity and the linear network topology are crucial in this work to make the cache-aided interference management tractable.

This network model is also highly correlated to the $K\times K$ partially connected \emph{circular} interference network where each receiver $i$ is connected to $L$ circulant transmitters denoted as $\mathcal{T}^c_{i}\triangleq[i:i+L-1]\bmod K$. Specifically, if we merge each of the first $L-1$ transmitters with each of the last $L-1$ transmitters, i.e., $j$ and $K+j$, for $j\in[0:L-2]$, in the linear network, then the merged node $(j, K+j)$ can be equivalent to transmitter $j$ in the circular network. This is because the connected receiver set of transmitter $j$ in the circular network, denoted as $\mathcal{R}_j^c \triangleq [j-L+1:j] \bmod K$, is the same as the union of the connected receiver sets of transmitters $j$ and $K+j$ in the linear network, i.e., $\mathcal{R}^c_{j}=\mathcal{R}_j\cup \mathcal{R}_{K+j}$. The original channel coefficients $h_{i,j}$ ($i\in \mathcal{R}_j$) and $h_{i,K+j}$ ($i\in\mathcal{R}_{K+j}$) can thus be viewed as the channel coefficients $h_{i,j}$ ($i\in\mathcal{R}^c_{j}$) in the circular network. Fig. \ref{Fig circular model} shows the resulting $4\times 4$ circular network by merging transmitter pairs $(0,4)$ and $(1,5)$ in Fig. \ref{Fig linear model} into transmitters 0 and 1, respectively.

While this paper mainly focuses on the constant receiver connectivity, we will also extend our results to the linear network with heterogeneous receiver connectivity, where each receiver $i$, for $i\in[0:K-1]$, is locally connected to $L_i$ consecutive transmitters $[i:i+L_i-1]$, in Section \ref{sec hete}. In this paper, we will first study the original linear network in Section \ref{section alignment} and \ref{section neutralization}, and then extend our analysis to the circular network and the heterogeneous linear network in Section \ref{sec circular}.

\subsection{Caching and Delivery Model}
Consider a database consisting of $N$ files ($N\ge K$), denoted by $\{W_0,W_1\cdots,W_{N-1}\}$, each of size $F$ bits. Each transmitter and receiver can cache at most $\mu_TNF$ bits and $\mu_RN F$ bits from the database, respectively, where $\mu_T\in[0,1]$ and $\mu_R\in[0,1]$ are the \textit{normalized cache sizes} at each transmitter and receiver, respectively.

The system operates in two phases, a \textit{cache placement phase} and a \textit{content delivery phase}. In the cache placement phase, each transmitter $j$ designs a caching function $\phi_{j}$ to map the database into its cached content $U_{j}$, where $U_j$ is a binary sequence of length no more than $\mu_TNF$ bits. Likewise, each receiver $i$ designs a caching function $\psi_{i}$ to map the database into a binary sequence of length no more than $\mu_RNF$ bits, denoted by $V_{i}$. The caching functions $\{\phi_{j}\}_{j=0}^{K+L-1},\{\psi_{i}\}_{i=0}^{K-1}$ are designed without the knowledge of future receiver demands, nor the instantaneous channel conditions in the delivery phase. But they need to know the network connection topology.  These caching functions in general can allow for arbitrary coding within or across the files. However, the proposed achievable schemes in this paper only consider file splitting without coding. Note that since each receiver is locally connected to $L$ transmitters, the cache sizes should satisfy $L\mu_TNF+\mu_RNF\!\ge\! NF$, or equivalently, $L\mu_T+\mu_R\ge1$, in order to meet the future receiver demands, similar to \cite{mine}.

In the delivery phase, each receiver $i$ requests a file $W_{d_i}$, for $d_i\in[0:N-1]$ \footnote{This paper focuses on the case where each receiver requests a single file  in each delivery phase. If each receiver requests multiple files  at the same time, we can deliver these files using multiple resource blocks, e.g., time slots, frequency sub-bands, or spatial sub-channels. Then, the proposed delivery schemes in this work still apply.}. We denote ${\bf d}\triangleq(d_i)^{K-1}_{i=0}$ as the demand vector. Each transmitter $j$ has an encoding function $\Lambda_j$ to map its cached content $U_j$, receiver demand ${\bf d}$, and network-wide channel realization $\mathbf{H}=[h_{i,j}(t)]_{i\in[0:K-1],j\in \mathcal{T}_i,t\in[T]}$ to the codeword $(X_j[t])^T_{t=1}\triangleq\Lambda_j(U_j,{\bf d},\mathbf{H})$, where $T$ is the length of the codeword. Each codeword $(X_j[t])_{t=1}^T$ has an average transmit power constraint $P$. Each receiver $i$ has a decoding function $\Gamma_i$ to decode $\hat{W}_{d_i}\triangleq\Gamma_i(V_i,(Y_i[t])^T_{t=1},\mathbf{H},{\bf d})$ of its desired file $W_{d_i}$ using its cached content $V_i$, received signal $(Y_i[t])^T_{t=1}$, channel realization $\mathbf{H}$, and demand $\mathbf{d}$.
The worst-case error probability is defined as $P_\epsilon\triangleq\max_\mathbf{d}\max_i\mathbb{P}(\hat{W}_{d_i}\ne W_{d_i})$. A caching and delivery scheme, consisting of a sequence of caching and coding functions $\mathcal{S}\triangleq\{\{\phi_{j}\}_{j=0}^{K+L-1},\{\psi_{i}\}_{i=0}^{K-1},\{\Lambda_j\}_{j=0}^{K+L-1},\{\Gamma_i\}_{i=0}^{K-1}\}$, indexed by file size $F$, is said to be feasible if $P_\epsilon\to 0$ when $F\to\infty$ for almost all channel realizations.

Throughout this paper, we focus on integer-point cache size pair $(\mu_T = p/L, \mu_R = q/L)$, where $p,q \in[0:L]$, to present our achievable caching and delivery schemes. These integer-point cache size pairs mean that the cumulative cache sizes at every set of $L$ transmitters and every set of $L$ receivers are both integers. The achievable scheme for the general cache size pair in the feasible region $L\mu_T + \mu_R \ge 1$ can be obtained by memory sharing between these integer-point cache size pairs. To avoid the trivial case with $\mu_R = 1$ so that each receiver can store the entire database, we exclude the case with $q=L$.

\subsection{Performance Metric}
Similar to \cite{simeone,mine,cao,gunduz,yuan,girgis2017decentralized,TaoInvited,gunduz_random,gunduz_partial}, we adopt normalized delivery time (NDT) as the performance metric.

\begin{definition}\label{def ndt}
The NDT for a given feasible caching and coding scheme is defined as
\begin{eqnarray}
\tau(\mu_T,\mu_R)\triangleq\lim_{P\to\infty}\lim_{F\to\infty}\sup\frac{\max\limits_{{\bf d}}T}{F/\log P}.\notag
\end{eqnarray}
Moreover, the optimal NDT is defined as $\tau^*(\mu_T,\mu_R)\triangleq\inf\{\tau(\mu_T,\mu_R):\tau(\mu_T,\mu_R)\rm \;is\;achievable\}$.
\end{definition}

\begin{remark}\label{remark tau}
The NDT is first defined in \cite{simeone} to capture the worst-case latency needed to serve any possible user demand $\mathbf{d}$, normalized by the required time to transmit a single file in a point-to-point baseline channel, in the high signal-to-noise ratio (SNR) regime. For a given feasible caching and delivery scheme, the NDT can be computed by counting the actual amount of information bits delivered to a receiver and the transmission rate to that receiver. Specifically, let $RF$ denote the number of information bits delivered to each receiver at a transmission rate of $d\cdot \log P +o(\log P)$  in the high SNR regime, where $d$ is the per-user DoF. Then, by Definition \ref{def ndt}, NDT can be computed as $\tau=R/d$.
\end{remark}

\section{Basic Achievable Scheme for Small Transmitter Cache Size}\label{section alignment}
In this section, we propose a novel caching and delivery scheme at $(\mu_T=1/L,\mu_R=q/L)$ with $q\in[0:L-1]$, where the accumulated cache size at the $L$ connected transmitters of each receiver is just enough to collectively store the entire database. Since this scheme is still applicable when $\mu_T>1/L$ by ignoring the redundant transmitter cache size, it is also referred to as the basic scheme in this paper. We will propose an enhanced scheme for $\mu_T>1/L$ to fully exploit transmitter cache and obtain a lower NDT in the next section.

\subsection{CC Strategy for $\mu_T = 1/L$}\label{sec alignment cache}
To take into account the partial connectivity, we propose a CC strategy with period of $L$. Specifically, we split each file $W_n$, for $n\in[0:N-1]$, into $\binom{L}{q}L$ equal-sized subfiles, denoted by $\{W_{n,\mathcal{Q}}^\zeta:\mathcal{Q}\subset[0:L-1],|\mathcal{Q}|=q,\zeta\in[0:L-1]\}$. Each subfile $W_{n,\mathcal{Q}}^\zeta$ is cached at receiver set $\{i:(i\bmod L)\in \mathcal{Q}\}$ and transmitter set $\{j:(j\bmod L)= \zeta\}$. By doing so, the cached contents at the transmitters with indices congruent modulo $L$ are repeated, and so are the cached contents at the receivers with indices congruent modulo $L$. The CC strategy ensures that each receiver can access the entire database through its connected $L$ consecutive transmitters.

By the CC strategy, each receiver $i$ caches $\binom{L-1}{q-1}L$ subfiles of each file $W_n$, given by $\{W_{n,\mathcal{Q}}^\zeta:\mathcal{Q}\subset[0:L-1],|\mathcal{Q}|=q,(i\bmod L)\in \mathcal{Q}, \zeta\in[0:L-1] \}$, and each transmitter $j$ caches $\binom{L}{q}$ subfiles of each file $W_n$, given by $\{W_{n,\mathcal{Q}}^\zeta:\mathcal{Q}\subset[0:L-1],|\mathcal{Q}|=q, \zeta=(j\bmod L)\}$. Since each subfile $W_{n,\mathcal{Q}}^{\zeta}$ has $\frac{F}{\binom{L}{q}L}$ bits, it is easy to verify that each receiver and each transmitter cache $\frac{q}{L}NF$ and $\frac{1}{L}NF$ bits in total, respectively, which satisfy the cache size constraints.

\subsection{Content Delivery Exploiting Coded Multicasting and Interference Alignment}\label{sec alignment delivery}
We consider the worst-case scenario where each receiver requests a distinct file. When two or more receivers request a same file, the proposed delivery strategy can still be applied by treating the requests as different files. Without loss of generality, we assume that receiver $i$ desires file $W_i$, for $i\in[0:K-1]$.  Given its local cache, receiver $i$ only needs subfiles:
\begin{align}
W_i^{\textrm{need}}\triangleq &\left\{W_{i,\mathcal{Q}}^\zeta:\mathcal{Q}\subset[0:L-1],|\mathcal{Q}|=q,Q\not\ni (i\bmod L),\zeta\in[0:L-1]\right\}. \notag
\end{align}
In the following, we introduce a virtual-receiver assisted delivery scheme that can exploit both coded multicasting and interference alignment. We first use the $6\times4$ network with $L=3$ (see Fig.~\ref{Fig linear model}) when $(\mu_T=\frac{1}{3},\mu_R=\frac{1}{3})$ as an example to present our delivery scheme in detail, and then proceed to the general algorithm.

\begin{example}\label{example linear alignment delivery}
($6\times4$ network with connectivity $L=3$ when $(\mu_T=\frac{1}{3},\mu_R=\frac{1}{3})$) By the CC strategy, each receiver already caches 3 subfiles of its requested file, and desires the rest 6 subfiles. We list these subfiles in Table \ref{table linear alignment delivery}.

\begin{table*}[tbp]
\centering
\caption{Subfiles to be delivered in the $6\times4$ network when $(\mu_T=\frac{1}{3},\mu_R=\frac{1}{3})$}
\label{table linear alignment delivery}
\begin{tabular}{|c|c|c|c|c|c|c|}
\hline
                & Cached at Tx 0            & Cached at Tx 1            & Cached at Tx 2            & Cached at Tx 3            & Cached at Tx 4            & Cached at Tx 5            \\ \hline
Desired by Rx 0 & $W_{0,\{1\}}^0,W_{0,\{2\}}^0$ & $W_{0,\{1\}}^1,W_{0,\{2\}}^1$ & $W_{0,\{1\}}^2,W_{0,\{2\}}^2$ & $\emptyset$               & $\emptyset$               & $\emptyset$               \\ \hline
Desired by Rx 1 & $\emptyset$               & $W_{1,\{0\}}^1,W_{1,\{2\}}^1$ & $W_{1,\{0\}}^2,W_{1,\{2\}}^2$ & $W_{1,\{0\}}^0,W_{1,\{2\}}^0$ & $\emptyset$               & $\emptyset$               \\ \hline
Desired by Rx 2 & $\emptyset$               & $\emptyset$               & $W_{2,\{0\}}^2,W_{2,\{1\}}^2$ & $W_{2,\{0\}}^0,W_{2,\{1\}}^0$ & $W_{2,\{0\}}^1,W_{2,\{1\}}^1$ & $\emptyset$               \\ \hline
Desired by Rx 3 & $\emptyset$               & $\emptyset$               & $\emptyset$               & $W_{3,\{1\}}^0,W_{3,\{2\}}^0$ & $W_{3,\{1\}}^1,W_{3,\{2\}}^1$ & $W_{3,\{1\}}^2,W_{3,\{2\}}^2$ \\ \hline
\end{tabular}
\end{table*}

We expand the network by introducing 4 virtual receivers $\{-2,-1,4,5\}$ as shown in Fig. \ref{Fig delivery}, so that each transmitter is connected to $L=3$ consecutive receivers. These virtual receivers have local caches and follow the same CC strategy as actual receivers. Specifically, receiver $-2$ and $4$ cache the same subfiles as in receiver $1$, since $-2\equiv 4\equiv1\pmod 3$, receiver $-1$ and $5$ cache the same subfiles as in receiver $2$, since $-1\equiv5\equiv2\pmod 3$. Note that these virtual receivers do not actually send any content request. But for notation convenience, we let each virtual receiver $i$, for $i\in\{-2, 1, 4, 5\}$, send a request to a virtual file $W_{i}$, with all-zero elements. By such network expansion, each subfile desired by an actual receiver is always cached in one (virtual) receiver that connects to the same transmitter as this actual receiver. Coded multicasting gain via XOR combining can thus be exploited among these two receivers. Specifically, transmitter $j$, for $j\in [0:5]$, can generate coded messages
\begin{align}
&\left\{W_{\textrm{Tx}_j}^{\{i_1,i_2\}}\triangleq W_{i_1,\{\hat{i}_2\}}^{\hat{j}}\oplus W_{i_2,\{\hat{i}_1\}}^{\hat{j}}: \forall i_1,i_2 \in[j-2:j],\{i_1,i_2\}\cap[0:3]\neq\emptyset\right\},\notag
\end{align}
where $\hat{j}=j\bmod L$, $\hat{i}_1=i_1\bmod L$, and $\hat{i}_2=i_2\bmod L$. Here, we use $W_{\textrm{Tx}_j}^{\{i_1,i_2\}}$ to denote the coded message sent by transmitter $j$ and needed by receiver group $\{i_1,i_2\}$ (if the receiver is a virtual one, it does not really need to decode the coded message.) Fig.~\ref{Fig delivery}  shows the flow of all coded messages in this example. Note that transmitter 0 and 5 only have two coded messages to send, because the virtual receiver groups $\{-2,-1\}$ and $\{4,5\}$ do not request any actual files.
\begin{figure}[!tbp]
\begin{centering}
\includegraphics[scale=0.15]{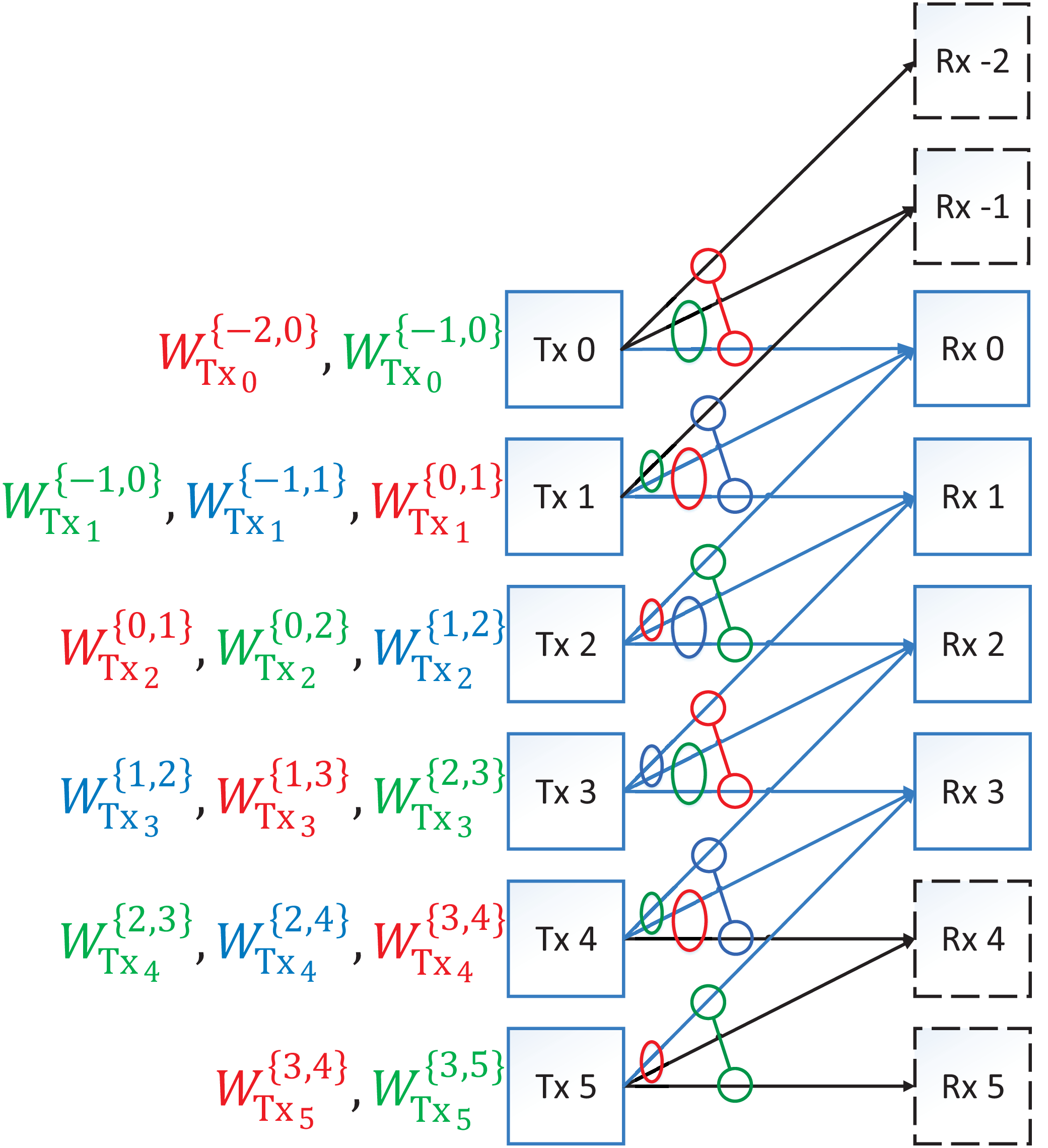}
\caption{Coded message flow of the subfile delivery in the $6\times 4$ network, where circles denote the multicast group of the message with the same color.}\label{Fig delivery}
\end{centering}
\end{figure}

By coded multicasting, the network is converted to an \emph{expanded} partially connected X-multicast channel with multicast size of 2, where every set of two receivers (should include at least one actual receiver) form a multicast group, and desire a common message from any of the transmitters connected to both of them. Each actual receiver desires six messages. Receiver 0 and 3 observe two undesired messages, and receiver 1 and 2 observe three undesired messages. We aim to adopt asymptotic interference alignment to align the undesired messages in a same subspace at each receiver. Specifically, we use a $T_n=6n^{6}+(n+1)^{6}$-symbol extension, with $n\in\mathbb{Z}^+$. Note that the symbol extension can be done in either time domain or frequency domain. We focus on the time domain in this paper, but our scheme is also directly applicable to the frequency domain if the channel coefficients are frequency-selective. Hence, the channel between transmitter $j$ and receiver $i$ ($j\in \mathcal{T}_i$) becomes a $T_n\times T_n$ diagonal matrix $\mathbf{H}_{i,j}$ whose diagonal entries $h_{i,j}(t)$ ($1\le t\le T_n$) are i.i.d. following some continuous distribution. We encode each message $W_{\textrm{Tx}_j}^{\{i_1,i_2\}}$ into a column vector of $n^6$ symbols $\mathbf{x}_{j}^{\{i_1,i_2\}}=[x_{j,m}^{\{i_1,i_2\}}]_{1\le m\le n^{6}}$, and  each symbol $x_{j,m}^{\{i_1,i_2\}}$ is beamformed along a $T_n\times1$ column vector $\mathbf{v}_{j,m}^{\{i_1,i_2\}}=\left[v_{j,m}^{\{i_1,i_2\}}(t)\right]_{1\le t\le T_n}$. Then, the codeword of message $W_{\textrm{Tx}_j}^{\{i_1,i_2\}}$ is given by $\mathbf{V}_{j}^{\{i_1,i_2\}}\mathbf{x}_{j}^{\{i_1,i_2\}}\triangleq\sum_{m}\mathbf{v}_{j,m}^{\{i_1,i_2\}}x_{j,m}^{\{i_1,i_2\}}$, where $\mathbf{V}_{j}^{\{i_1,i_2\}}$ is a $T_n\times n^6$ matrix. We need to design $\{\mathbf{V}_{j}^{\{i_1,i_2\}}\}$ to satisfy the interference alignment conditions
\begin{equation}
\left\{\!\!
\begin{array}{ll}
  \textrm{Rx 0:} &\mathbf{H}_{0,1}\mathbf{V}_{1}^{\{-1,1\}}\!\prec\!\mathbf{I}_0,\ \mathbf{H}_{0,2}\mathbf{V}_{2}^{\{1,2\}}\!\prec\!\mathbf{I}_0, \\
  \textrm{Rx 1:} &\mathbf{H}_{1,1}\mathbf{V}_{1}^{\{-1,0\}}\!\prec\!\mathbf{I}_1,\ \mathbf{H}_{1,2}\mathbf{V}_{2}^{\{0,2\}}\!\prec\!\mathbf{I}_1,\  \mathbf{H}_{1,3}\mathbf{V}_{3}^{\{2,3\}}\!\prec\!\mathbf{I}_1,\\
  \textrm{Rx 2:} &\mathbf{H}_{2,2}\mathbf{V}_{2}^{\{0,1\}}\!\prec\!\mathbf{I}_2,\ \;\,\,\mathbf{H}_{2,3}\mathbf{V}_{3}^{\{1,3\}}\!\prec\!\mathbf{I}_2,\ \mathbf{H}_{2,4}\mathbf{V}_{4}^{\{3,4\}}\!\prec\!\mathbf{I}_2,\\
  \textrm{Rx 3:} &\mathbf{H}_{3,3}\mathbf{V}_{3}^{\{1,2\}}\!\prec\!\mathbf{I}_3,\ \;\,\,\mathbf{H}_{3,4}\mathbf{V}_{4}^{\{2,4\}}\!\prec\!\mathbf{I}_3,
\end{array}
\right.\label{eqn example alignment}
\end{equation}
where $\mathbf{I}_0,\mathbf{I}_1,\mathbf{I}_2,\mathbf{I}_3$ are four $T_n\times (n+1)^6$ matrices. Condition \eqref{eqn example alignment} implies that, for each receiver, the received beamforming vectors of all its undesired messages are aligned together in a subspace with $(n+1)^6$ dimensions. For example at receiver 0, the received beamforming vectors of $W_{\textrm{Tx}_1}^{\{-1,1\}}$ and $W_{\textrm{Tx}_2}^{\{1,2\}}$, given by the column vectors of $\mathbf{H}_{0,1}\mathbf{V}_{1}^{\{-1,1\}}$ and  $\mathbf{H}_{0,2}\mathbf{V}_{2}^{\{1,2\}}$ respectively, are aligned together in the space spanned by the $(n+1)^6$ column vectors of $\mathbf{I}_0$. The detailed design of  matrices $\{\mathbf{V}_{j}^{\{i_1,i_2\}}\}$ and $\{\mathbf{I}_i:i\in[0:3]\}$ are given in Appendix A. Note that the design of the entries in the $t$-th row of $\{\mathbf{V}_{j}^{\{i_1,i_2\}}\}$ in Appendix A only needs the current channel realization $\{h_{i,j}(t):i\in[0:K-1],j\in\mathcal{T}_i\}$ at time slot $t$, for $t\in[1:T_n]$, since the channel matrices $\{\mathbf{H}_{i,j}\}$ are all diagonal. Thus, the required CSI for the design of transmit beamforming vectors is causal. Given that each transmitter has $n^{6}$ beamforming vectors of each of its transmitted messages, and each beamforming vector has $T_n$ entries, the beamforming vector design for transmitters in this example has a computational complexity of  $O(n^{12})$.

By using this scheme, Fig.~\ref{Fig IA} shows the signal space of receiver 0 as an example. After receiving all signals in $T_n$-symbol extension, the undesired messages are aligned together at $(n+1)^6$ dimensions, and each receiver can decode its six desired messages, each taking up $n^6$ dimensions, by constructing a $T_n\times T_n$ receive combining matrix with computational complexity of $O(n^{18})$, as given in Appendix A. The required CSI for the design of receive combining matrices is also causal. Hence, a per-user DoF of $\frac{6n^6}{6n^6+(n+1)^6}$ is achieved. By letting $n\rightarrow\infty$, a per-user DoF of $\frac{6}{7}$ is achieved.

\begin{figure}[!tbp]
\begin{centering}
\includegraphics[scale=0.2]{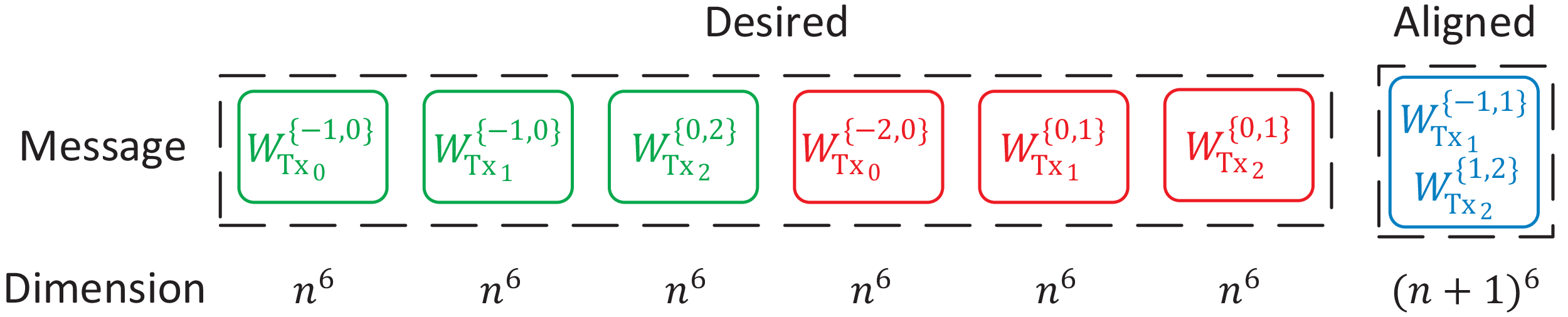}
\caption{Signal space of receiver 0 in the $6\times 4$ network.}\label{Fig IA}
\end{centering}
\end{figure}

Since each receiver desires six messages, each with $\frac{1}{9}F$ bits, the achievable NDT in the $6\times 4$ linear network when $(\mu_T=\frac{1}{3},\mu_R=\frac{1}{3})$ can be computed, by Remark \ref{remark tau}, as $\tau=\frac{6\times 1/9}{6/7}=\frac{7}{9}$.
\end{example}

Now, we proceed to consider the delivery strategy in the general $(K+L-1)\times K$ partially connected linear network. Following Example \ref{example linear alignment delivery}, to ensure that each transmitter is connected to $L$ receivers, we introduce $2L-2$ virtual receivers to transform the network into the \emph{expanded} partially connected linear interference network. Each virtual receiver adopts the same CC strategy as an actual receiver, but does not send any content request. In the expanded network, each subfile desired by one actual receiver is cached in $q$ (virtual) receivers connecting to the same transmitter as this actual receiver. Coded multicasting gain via XOR combining can thus be exploited among these $q+1$ receivers. The detailed generation of the expanded network and coded multicast messages is shown in Algorithm \ref{algorithm 1}.

\begin{algorithm}[t]
\caption{Network expansion and coded message generation in the $(K+L-1)\times K$ partially connected linear interference network}\label{algorithm 1}
\begin{algorithmic}[1]
\STATE Generate virtual receivers $[-L+1:-1]\cup[K:K+L-2]$, each virtual receiver $i$ is connected to transmitters $[i:i+L-1]\cap[0:K+L-2]$. Denote $\mathcal{R}_j^e\triangleq[j-L+1:j]$ as the set of (virtual) receivers connected to transmitter $j$
\STATE Virtual receiver $i\in[-L+1:-1]$ caches the same subfiles as in actual receiver $i+L$, and virtual receiver $i\in[K:K+L-2]$ caches the same subfiles as in actual receiver $i-L$
\FOR{$j=0,1,\ldots,K+L-2$}
\FOR{$\mathcal{R}\subseteq\mathcal{R}_j^e$, $|\mathcal{R}|=q+1$}
\IF{$\mathcal{R}\cap[0:K-1]\neq\emptyset$} \STATE Transmitter $j$ generates coded message $W_{\textrm{Tx}_j}^{\mathcal{R}}=\bigoplus\limits_{i\in\mathcal{R}\cap[0:K-1]}W_{i,{\hat{\mathcal{R}}\backslash\{\hat{i}\}}}^{\hat{j}}$, where $\hat{j}=j\bmod L$, $\hat{i}=i\bmod L$, $\hat{\mathcal{R}}=\{\hat{i}:i\in\mathcal{R}\}$.\ENDIF
\ENDFOR
\ENDFOR
\end{algorithmic}
\end{algorithm}

Based on Algorithm \ref{algorithm 1}, the network topology is converted into the \emph{expanded} partially connected X-multicast channel with multicast size $q+1$,  where each transmitter $j$ has an independent coded message $W_{\textrm{Tx}_j}^\mathcal{R}$ intended for the actual receivers in multicast group $\mathcal{R}$ satisfying $\mathcal{R}\subseteq\mathcal{R}_j^e,\mathcal{R}\cap[0\!:\!K\!-\!1]\!\neq\!\emptyset$, and $|\mathcal{R}|\!=\!q\!+\!1$. By using a novel scheme based on asymptotic interference alignment, the DoF of this channel is given in the following lemma, whose proof is in Appendix A.

\begin{lemma}\label{lemma alignment dof}
The achievable per-user DoF of the expanded $(K+L-1)\times K$ partially connected X-multicast channel with receiver connectivity $L$ and multicast size $q+1$ is
\begin{align}
d=\frac{L}{L-1+\frac{L}{q+1}}.\label{eqn lemma1}
\end{align}
\end{lemma}

\emph{Sketch of the proof}: The main idea of the achievable scheme is first to divide all multicast messages into $\binom{L}{q+1}$ sets according to their intended receiver multicast groups. Then design beamforming vectors of each set to align the messages from this set in a same subspace at each undesired receiver by using asymptotic interference alignment.


The per-user DoF in \eqref{eqn lemma1} appears the same as the per-user DoF of an $L\times L$ fully connected X-multicast channel with muticast size $q+1$ in \cite[Theorem 2]{niesen}. This is because by our proposed interference alignment technique, each actual receiver in the expanded partially connected X-multicast channel sees an equivalent $L\times L$ fully connected X-multicast channel, independent of $K$. Our proof is similar to \cite{niesen}, but the message grouping for alignment is different. In our work, the messages in each set are intended for different receivers and not all transmitters have messages to send in each set, while in \cite{niesen}, all the messages in each set are intended for the same receivers and each transmitter has exactly one message in each set.

\begin{remark}\label{reamrk virtual receiver}
The purpose of introducing virtual receivers during the delivery phase is two-fold. The first is to unify the notation of coded messages in line 6 of Algorithm \ref{algorithm 1}. The second is to convert the channel during the delivery phase into an expanded partially connected X-multicast channel, for which the DoF analysis is more tractable.  The virtual receivers however do not affect the actual DoF results of the original unexpanded network, since they do not send any content request, neither intend to decode any message.
\end{remark}

Since each receiver desires $\binom{L-1}{q}L$ subfiles, each with $\frac{1}{\binom{L}{q}L}F$ bits, then, by Remark \ref{remark tau} and Lemma \ref{lemma alignment dof}, the achievable NDT at integer-point cache size pair $(\mu_T=\frac{1}{L},\mu_R=\frac{q}{L})$ is
\begin{align}
\tau(\mu_T=\frac{1}{L},\mu_R=\frac{q}{L})&=\frac{\binom{L-1}{q}L\times\frac{1}{\binom{L}{q}L}}{d}=\frac{\left(L-1+\frac{L}{q+1}\right)\left(L-q\right)}{L^2}.\label{eqn artime}
\end{align}

\subsection{Achievable NDT and Discussions}
As mentioned before, the proposed basic scheme is also applicable when $\mu_T\ge 1/L$. Then, the achievable NDT for the entire cache size region in the partially connected linear interference network is formally given in the following theorem.

\begin{theorem}\label{thm achievable NDT alignment}
(Achievable NDT-1 for linear network) For the cache-aided $(K+L-1)\times K$ partially connected linear interference network, an achievable NDT-1 is given by
\begin{align}
\tau_{\textrm{A1}}(\mu_T\ge\frac{1}{L},\mu_R=\frac{q}{L})\triangleq& \frac{\left(L-1+\frac{L}{q+1}\right)\left(L-q\right)}{L^2}, \quad \textrm{ if } q\in[0:L-1].\label{eqn thm 1}
\end{align}
For general feasible cache size pair $(\mu_T,\mu_R)$ satisfying $L\mu_T+\mu_R\ge1$, the achievable NDT-1 is given by the lower convex envelope of these points  and the trivial points $\tau(\mu_T\ge0,\mu_R=1)=0$ using memory sharing.
\end{theorem}

To better understand the behavior of this scheme, we decompose \eqref{eqn thm 1} into two components:
\begin{align}
\tau_\textrm{A1}&=\frac{L-1+\frac{L}{q+1}}{L}\cdot\frac{L-q}{L}=\left(1-\frac{1}{L}+\frac{1}{1+L\mu_R}\right)\cdot\left(1-\mu_R\right).\label{eqn integer tau 1}
\end{align}
The expression in \eqref{eqn integer tau 1} suggests that this scheme exploits the receiver local caching gain of $(1-\mu_R)$ and a combined coded multicasting and interference alignment gain of $(1-\frac{1}{L}+\frac{1}{1+L\mu_R})$.

Though the proposed basic scheme focuses on $\mu_T=\frac{1}{L}$ and ignores the redundant transmitter cache size when $\mu_T>\frac{1}{L}$, the achievable NDT-1 in Theorem \ref{thm achievable NDT alignment} is still within a constant multiplicative gap to the optimum in the entire cache size region, and even  optimal in certain cache size regions. These are formally stated in the following theorems.

\begin{theorem}\label{thm gap}
(Multiplicative Gap) For the considered network, the multiplicative gap between the achievable NDT-1 and the optimal NDT is less than 2, i.e., $\frac{\tau_{\textrm{A1}}}{\tau^*}<2$.
\end{theorem}

\begin{theorem}\label{thm optimal NDT alignment}
(Optimality) For the considered network, the achievable NDT-1 is optimal if $\mu_R\ge\max\{1-\mu_T,\frac{L-1}{L}\}$, with $\tau^*=1-\mu_R$.
\end{theorem}

The proof of Theorem \ref{thm gap} is given in Appendix B, while the proof of Theorem \ref{thm optimal NDT alignment} is similar to Appendix B, by comparing the achievable NDT-1 using memory sharing between $(\mu_T=\frac{1}{L},\mu_R=\frac{L-1}{L})$ and $(\mu_T=0,\mu_R=1)$ to the lower bound \eqref{eqn lowerbound 1} in Appendix B, and thus neglected. There is an intuitive explanation for Theorem \ref{thm optimal NDT alignment}. At $(\mu_T=\frac{1}{L},\mu_R=\frac{L-1}{L})$, each subfile desired by one actual receiver is cached at $L-1$ (virtual) receivers connected to the same transmitter as this receiver, and each transmitter thus can use bit-wise XOR to generate a coded message intended for all of its connected actual receivers. As a result, each actual receiver sees an equivalent ideal single-user interference-free channel, and can obtain an optimal NDT of $\tau^*=\frac{1}{L}$. At $(\mu_T=0,\mu_R=1)$, each receiver can cache the entire database, yielding the optimal NDT of $\tau^*=0$. Thus, by using memory sharing between these two cache size pairs, each receiver still sees an ideal single-user interference-free channel, and can obtain the optimal NDT of $\tau^*=1-\mu_R$.

\begin{figure}[!t]
\begin{centering}
\includegraphics[scale=0.33]{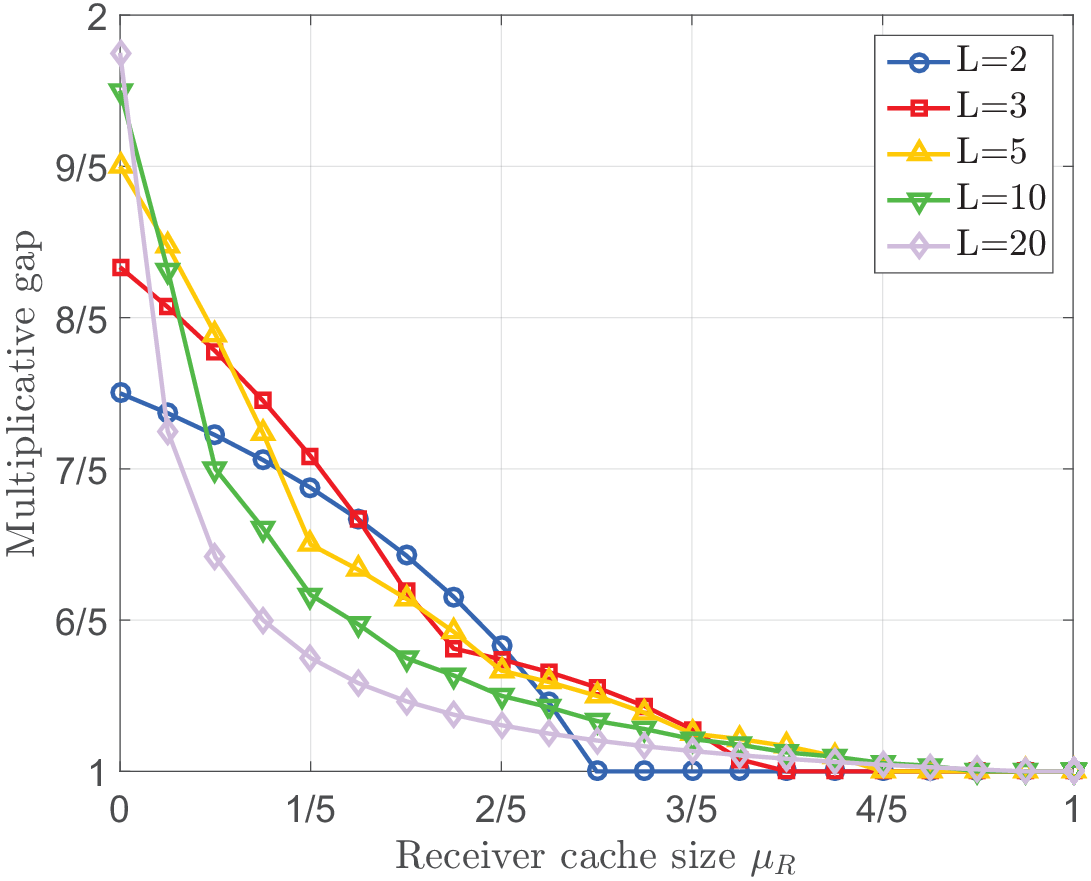}
\caption{Multiplicative gap between the achievable NDT in Theorem \ref{thm achievable NDT alignment} and the lower bound \eqref{eqn lowerbound 1} in Appendix B.}\label{Fig GapAlignment}
\end{centering}
\vspace{-15pt}
\end{figure}

\begin{remark}[Impact of transmitter cooperation]
In the basic scheme, each transmitter only caches $\frac{1}{L}NF$ bits and does not fully utilize its cache storage when $\mu_T>\frac{1}{L}$. Therefore, transmitter cooperation is not fully exploited. Hence, Theorem \ref{thm gap} implies that transmitter cooperation cannot provide any gain more than a constant factor of 2, similar to \cite[Section III-1]{niesen} in the fully connected networks. Fig. \ref{Fig GapAlignment} plots the multiplicative gap between the achievable NDT-1 and the lower bound \eqref{eqn lowerbound 1} in Appendix B. If $\mu_R\!\ge\!\max\{1\!-\!\mu_T,\!\frac{L-1}{L}\}$, the gap is 1 since the achievable NDT-1 is optimal as shown in Theorem \ref{thm optimal NDT alignment}. If $\mu_R\!<\!\max\{1\!-\!\mu_T,\!\frac{L-1}{L}\}$, the  gap is a decreasing function of $\mu_R$, and is close to 1 when $\mu_R$ is large, e.g., it is less than $\frac{6}{5}$ when $(\mu_T\!\ge\!\frac{1}{L},\mu_R\!\ge\!\frac{2}{5})$. This is because when $\mu_R$ is large, by using the interference alignment technique in Appendix A, most dimensions of the received signal space at each receiver are occupied by its desired messages, hence the per-user DoF is close to 1. Thus, the achievable NDT approaches the lower bound \eqref{eqn lowerbound 1}, and the multiplicative gap is close to 1. Fig. \ref{Fig GapAlignment} further implies that the reduction of NDT is very limited by exploiting transmitter cooperation when $\mu_R$ is large.
\end{remark}

\begin{remark}[Computational complexity of the basic scheme]\label{remark complexity basic}
In the cache placement phase, the computational complexity mainly comes from the file partition overhead, and partitioning one file into $n$ subfiles has a complexity of $O(n)$. Since we split each file into $L\binom{L}{q}$ subfiles, the computational complexity of our CC strategy is given by $O(NL^{q+1})$. In the content delivery phase, the complexity mainly comes from  the transmit beamforming vector design for transmitters and the receive combining matrix design for receivers. As shown in Appendix A, each transmitter needs $n^{(K+L-1)(L-q-1)}$ beamforming vectors of each of its transmitted messages in $T_n$-symbol extension  with $n\in\mathbb{Z}^+$ and $T_n$ defined in Appendix A, and the beamforming vectors are given in \eqref{eqn appendix a matrix set}. The complexity of the transmit beamforming vector design is given by $O(K^2L^{2q+3}n^{2(K+L-1)(L-q-1)})$. After receiving all signals in $T_n$-symbol extension, the receive combining matrix for each receiver $i$ is given by $\mathbf{A}_i^{-1}$, where $\mathbf{A}_i$ is the $T_n\times T_n$ matrix defined in Appendix A. Thus, the complexity of the receive combining matrix design comes from the inverse operation of $\mathbf{A}_i$ for all $i\in[0:K-1]$, and is given by $O(KL^{3q+3}n^{3(K+L-1)(L-q-1)})$.
\end{remark}

\section{Enhanced Achievable Scheme for Large Transmitter Cache Size}\label{section neutralization}
In this section, we present an enhanced caching and delivery scheme at $(\mu_T=p/L,\mu_R=q/L)$ with $p\in[2:L],q\in[0:L-1]$ to fully utilize the transmitter cache and exploit interference neutralization gain opportunistically through transmitter cooperation in the delivery phase. Then, by combining this scheme and the basic scheme in Section \ref{section alignment}, we obtain a lower achievable NDT which is exactly optimal in a larger cache size region.
\subsection{Modified CC Strategy for $\mu_T = p/L$ with $p\ge2$}\label{sec neutralization cache}
At integer point cache size pair $(\mu_T=p/L,\mu_R=q/L)$ with $p\in[2:L],q\in[0:L-1]$, following Section \ref{sec alignment cache}, a natural way is to split each file into $\binom{L}{p}\binom{L}{q}$ subfiles, each cached at a distinct set of $p$ transmitters and $q$ receivers out of any $L$ consecutive transmitters and receivers, respectively.  However, this may bring significant challenge for the interference management in the delivery phase. Instead, we propose a modified CC strategy that only splits each file into $\binom{L}{q}(L-q)$ subfiles and consequently considerably reduces the complexity in the delivery phase.

More specifically, we first split each file $W_n$, for $n\in[0:N-1]$, into $\binom{L}{q}$ equal-sized subfiles, denoted by $\{W_{n,\mathcal{Q}}:\mathcal{Q}\subset[0:L-1],|\mathcal{Q}|=q\}$. Each subfile $W_{n,\mathcal{Q}}$ is cached at receiver set $\{i:(i\bmod L)\in \mathcal{Q}\}$. For each $\mathcal{Q}$, define its complement set $\bar{\mathcal{Q}}\triangleq[0:L-1]\backslash \mathcal{Q}=\{\zeta_1,\zeta_2,\ldots,\zeta_{L-q}\}$, where $\zeta_1<\zeta_2<\cdots<\zeta_{L-q}$. Then we further split each subfile $W_{n,\mathcal{Q}}$ into $L-q$ equal-sized subfiles, denoted by $\{W_{n,\mathcal{Q}}^{\zeta_1},W_{n,\mathcal{Q}}^{\zeta_2},\ldots,W_{n,\mathcal{Q}}^{\zeta_{L-q}}\}$, and cache each $W_{n,\mathcal{Q}}^{\zeta_u}$, for $u\in[L-q]$, at transmitter set $\{j:(j\bmod L)\in[\zeta_u+L-p:\zeta_u+L-1]\bmod L\}$. This modified CC strategy is similar to the one in Section \ref{sec alignment cache} at receivers in that every $q$ out of any $L$ consecutive receivers cache common subfiles of each file, but differs at transmitters and cannot degenerate to the latter even when $p=1$. Here, each subfile is repeatedly cached at $p$ consecutive transmitters in order to enable transmitter cooperation in the delivery phase.

By the modified CC strategy, each receiver $i$ caches $\binom{L-1}{q-1}(L-q)$ subfiles of each file $W_n$, given by $\{W_{n,\mathcal{Q}}^{\zeta_u}:\mathcal{Q}\subset[0:L-1],|\mathcal{Q}|=q,\mathcal{Q}\ni (i\bmod L), \zeta_u\in[0:L-1]\backslash \mathcal{Q}\}$, and each transmitter $j$ caches $\binom{L-1}{q}p$ subfiles of each file $W_n$, given by $\{W_{n,\mathcal{Q}}^{\zeta_u}:\mathcal{Q}\subset[0:L-1],|\mathcal{Q}|=q,\zeta_u\in[0:L-1]\backslash \mathcal{Q},\zeta_u\in[j-L+1:j-L+p]\bmod L\}$. Since each subfile has $\frac{F}{\binom{L}{q}(L-q)}$ bits, it is easy to verify that each receiver and each transmitter cache $\frac{q}{L}NF$ and $\frac{p}{L}NF$ bits in total, respectively, meaning that all the caches are fully filled.

\subsection{Content Delivery Exploiting Self-Interference Cancellation and Interference Neutralization}\label{sec neutralization delivery}
Similar to Section \ref{sec alignment delivery}, we still assume that receiver $i$ desires file $W_i$, for $i\in[0:K-1]$, without loss of generality. Given local caches, each receiver $i$ only needs subfiles
\begin{align}
W_i^{\textrm{need}}\triangleq&\left\{W_{i,\mathcal{Q}}^{\zeta_u}:\mathcal{Q}\subset[0:L-1],|\mathcal{Q}|=q,\mathcal{Q}\not\ni (i\bmod L),\zeta_u\in[0:L-1]\backslash \mathcal{Q}\right\}.\notag
\end{align}
The delivery scheme in this section differs from that in Section \ref{sec alignment delivery}. It does not need virtual receivers, and it can exploit interference neutralization rather than interference alignment by making use of the overlapped cache contents at neighboring transmitters. In the following, we first use the $6\times4$ network when $(\mu_T=\frac{2}{3},\mu_R=\frac{1}{3})$ as an example to present our delivery scheme, and then proceed to the general scheme.
\begin{example}\label{example linear neutralization delivery}
($6\times4$ network with connectivity $L=3$ when $(\mu_T=\frac{2}{3},\mu_R=\frac{1}{3})$) By the modified CC strategy, each receiver already caches 2 subfiles, and only needs the rest 4 subfiles. We list these subfiles in Table \ref{table linear neutralization delivery}.

\begin{table*}[tbp]
\centering
\caption{Subfiles to be delivered in the $6\times4$ network when $(\mu_T=\frac{2}{3},\mu_R=\frac{1}{3})$}
\label{table linear neutralization delivery}
\begin{tabular}{|c|c|c|c|}
\hline
 & Cached at Tx set $\{0,1,3,4\}$ & Cached at Tx set $\{0,2,3,5\}$ & Cached at Tx set $\{1,2,4,5\}$ \\ \hline
Desired by Rx 0 & $W_{0,\{1\}}^2$ & $W_{0,\{2\}}^1$ & $W_{0,\{1\}}^0,W_{0,\{2\}}^0$ \\ \hline
Desired by Rx 1 & $W_{1,\{0\}}^2$ & $W_{1,\{0\}}^1,W_{1,\{2\}}^1$ & $W_{1,\{2\}}^0$ \\ \hline
Desired by Rx 2 & $W_{2,\{0\}}^2,W_{2,\{1\}}^2$ & $W_{2,\{0\}}^1$ & $W_{2,\{1\}}^0$ \\ \hline
Desired by Rx 3 & $W_{3,\{1\}}^2$ & $W_{3,\{2\}}^1$ & $W_{3,\{1\}}^0,W_{3,\{2\}}^0$ \\ \hline
\end{tabular}
\end{table*}


We use a 4-symbol extension to transmit subfiles in Table \ref{table linear neutralization delivery}. Denote $x_{i,\mathcal{Q}}^{\zeta_u}$ as the transmitted symbol of subfile $W_{i,\mathcal{Q}}^{\zeta_u}$. At each time slot $t$, denote $v_{i,\mathcal{Q}}^{\zeta_u,j}(t)$ as the precoding coefficient of $x_{i,\mathcal{Q}}^{\zeta_u}$ at transmitter $j$ which caches $W_{i,\mathcal{Q}}^{\zeta_u}$, and column vector $\mathbf{v}_{i,\mathcal{Q}}^{\zeta_u}(t)$ as the $6\times1$ cooperative beamforming vector of $x_{i,\mathcal{Q}}^{\zeta_u}$ by combining the precoding coefficients from all the 6 transmitters together. Note that for those transmitters not caching $W_{i,\mathcal{Q}}^{\zeta_u}$, their corresponding entries in $\mathbf{v}_{i,\mathcal{Q}}^{\zeta_u}(t)$ are 0. For example, the beamforming vector of $x_{2,\{1\}}^{0}$ at each time slot $t$ is $\mathbf{v}_{2,\{1\}}^{0}(t)\!=\!\big[0,v_{2,\{1\}}^{0,1}(t),v_{2,\{1\}}^{0,2}(t),0,v_{2,\{1\}}^{0,4}(t),v_{2,\{1\}}^{0,5}(t)\big]\!^\top$. Then, the received signals at 4 receivers at each time slot $t$, for $t\in[4]$, are given by
\begin{equation}
\left[
\begin{array}{c}
  y_0(t) \\
  y_1(t) \\
  y_2(t) \\
  y_3(t)
\end{array}
\right]=
\left[
\begin{array}{cccccc}
  h_{0,0}(t) & h_{0,1}(t) & h_{0,2}(t) & 0 & 0 & 0 \\
  0 & h_{1,1}(t) & h_{1,2}(t) & h_{1,3}(t) & 0 & 0 \\
  0 & 0 & h_{2,2}(t) & h_{2,3}(t) & h_{2,4}(t) & 0 \\
  0 & 0 & 0 & h_{3,3}(t) & h_{3,4}(t) & h_{3,5}(t)
\end{array}
\right]\cdot\sum_{i,\mathcal{Q},\zeta_u} \mathbf{v}_{i,\mathcal{Q}}^{\zeta_u}(t)x_{i,\mathcal{Q}}^{\zeta_u}.\label{eqn example neutralization channel}
\end{equation}
The goal is to design the beamforming vector of each symbol so as to eliminate interference at undesired receivers at all 4 time slots. Here, we use $\mathbf{v}_{2,\{1\}}^0(t)$ as an example to illustrate the generation of these vectors. Since $x_{2,\{1\}}^0$ is desired by receiver 2 and unwanted by receivers 0,1,3, the precoding coefficients in $\mathbf{v}_{2,\{1\}}^{0}(t)$ should satisfy
\begin{equation}
\left\{
\begin{array}{ll}
h_{2,2}(t)v_{2,\{1\}}^{0,2}(t)+h_{2,4}(t)v_{2,\{1\}}^{0,4}(t)\neq0 &\textrm{ at receiver 2},\\
h_{0,1}(t)v_{2,\{1\}}^{0,1}(t)+h_{0,2}(t)v_{2,\{1\}}^{0,2}(t)=0  &\textrm{ at receiver 0},\\
h_{3,4}(t)v_{2,\{1\}}^{0,4}(t)+h_{3,5}(t)v_{2,\{1\}}^{0,5}(t)=0  &\textrm{ at receiver 3},
\end{array}
\right.\label{eqn example neutralization}
\end{equation}
for $t\in[4]$. Note that receiver 1 can cancel $x_{2,\{1\}}^0$ directly using its local cache since it has cached it. One solution of \eqref{eqn example neutralization} is $\mathbf{v}_{2,\{1\}}^{0}(t)\!=\!\alpha_{2,\{1\}}^{0}(t)\!\cdot\!\big[0,-\frac{h_{0,2}(t)}{h_{0,1}(t)},1,0,0,0\big]\!^\top$ at each time slot $t$, where $\alpha_{2,\{1\}}^{0}(t)$ follows some continuous distribution. Note that transmitter 1, though not connected to receiver 2, still needs to transmit $x_{2,\{1\}}^0$ so as to neutralize it at receiver 0. In general, the transmission of subfiles desired by receivers not connected to that transmitter is a key method of the enhanced scheme and a key difference from the basic scheme and \cite{Wynercaching}. By designing $\mathbf{v}_{2,\{1\}}^{0}(t)$, symbol $x_{2,\{1\}}^0$ can be received by receiver 2 without causing interference to other receivers. Similarly, we can design the beamforming vectors of other symbols to eliminate interferences at unwanted receivers, and each receiver only receives its desired symbols at all 4 time slots. An interference-free transmission is thus created for all receivers. Fig. \ref{Fig IN} shows the received subfiles and neutralized subfiles of receiver 2 as an example. By canceling cached subfiles  using local cache, the signal space of receiver 2 is only occupied by its 4 desired subfiles which can be decoded in 4-symbol extension. Thus, a per-user DoF of 1 is achieved.

\begin{figure*}[!tbp]
\begin{centering}
\includegraphics[scale=0.35]{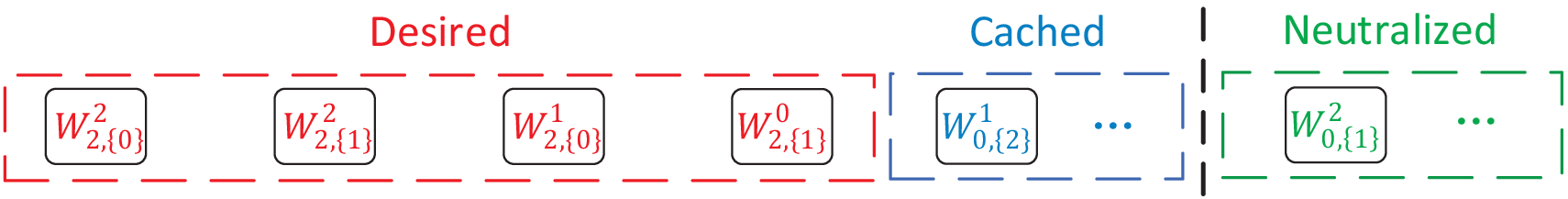}
\caption{Received subfiles and neutralized subfiles of receiver 2 in the $6\times 4$ network.}\label{Fig IN}
\end{centering}
\vspace{-10pt}
\end{figure*}

Since each receiver desires 4 subfiles, each with $\frac{F}{6}$ bits, the achievable NDT is given by $\tau=\frac{4\times1/6}{1}=\frac{2}{3}$.
\end{example}

Now we proceed to the general $(K+L-1)\times K$ partially connected linear network. We first consider the case when $p+q\ge L$. We use a $\binom{L-1}{q}(L-q)$-symbol extension to transmit subfiles, where $\binom{L-1}{q}(L-q)$ is also the number of subfiles desired by each receiver. Similar to Example \ref{example linear neutralization delivery}, denote $x_{i,\mathcal{Q}}^{\zeta_u}$ as the transmitted symbol of subfile $W_{i,\mathcal{Q}}^{\zeta_u}$. At each time slot $t$, denote $v_{i,\mathcal{Q}}^{\zeta_u,j}(t)$ as the precoding coefficient of $x_{i,\mathcal{Q}}^{\zeta_u}$ at transmitter $j$ which caches $W_{i,\mathcal{Q}}^{\zeta_u}$, and column vector $\mathbf{v}_{i,\mathcal{Q}}^{\zeta_u}(t)$ as the $(K+L-1)\times 1$ cooperative beamforming vector of $x_{i,\mathcal{Q}}^{\zeta_u}$ by combining the precoding coefficients from all the $K+L-1$ transmitters together. Note that for those transmitters not caching $W_{i,\mathcal{Q}}^{\zeta_u}$, their corresponding entries in $\mathbf{v}_{i,\mathcal{Q}}^{\zeta_u}(t)$ are 0. At each time slot $t$, for $t\in[\binom{L-1}{q}(L-q)]$, the received signals at all receivers are given by
\begin{align}
  \mathbf{y}(t)=\mathbf{H}(t)\cdot\sum_{i,\mathcal{Q},\zeta_u} \mathbf{v}_{i,\mathcal{Q}}^{\zeta_u}(t)x_{i,\mathcal{Q}}^{\zeta_u},\notag
\end{align}
where $\mathbf{y}(t)\in\mathbb{C}^{K\times1}$ denotes the received signals at all receivers, and $\mathbf{H}(t)$ is the $K\times (K+L-1)$ channel matrix between all transmitters and receivers. Note that each symbol $x_{i,\mathcal{Q}}^{\zeta_u}$ is desired by receiver $i$, and unwanted by the rest $K-1$ receivers. Among these $K-1$ receivers, $q$ out of any $L$ consecutive receivers have cached $W_{i,\mathcal{Q}}^{\zeta_u}$, and can cancel $x_{i,\mathcal{Q}}^{\zeta_u}$ directly. Thus, we only need to design the beamforming vector $\mathbf{v}_{i,\mathcal{Q}}^{\zeta_u}(t)$ to neutralize $x_{i,\mathcal{Q}}^{\zeta_u}$ at the rest unwanted receivers, and guarantee that $x_{i,\mathcal{Q}}^{\zeta_u}$ can be received and decoded successfully by receiver $i$. The detailed design of the vectors is given in Appendix C. The main idea is that, for each $x_{i,\mathcal{Q}}^{\zeta_u}$, we first design the precoding coefficients at $p$ consecutive cached transmitters $[j:j+p-1]$ ($\equiv [\zeta_u+L-p:\zeta_u+L-1]\pmod L$) with $j+p-1\in[i:i+L-1]$, to eliminate interference at $L-1$ unwanted receivers $[j+p-L:j+p-1]\backslash\{i\}$ near receiver $i$.
Note that $x_{i,\mathcal{Q}}^{\zeta_u}$ still causes interference to receivers $[j-L+1:j+p-L-1]$, we then design the precoding coefficients at the previous $p$ consecutive cached transmitters $[j-L:j+p-L-1]$ to neutralize $x_{i,\mathcal{Q}}^{\zeta_u}$ at them. Iteratively, each symbol will be neutralized at all unwanted and uncached receivers. Thus, each receiver only receives its $\binom{L-1}{q}(L-q)$ desired symbols, and an interference-free transmission is created for all receivers. Also by designing an i.i.d. multiplicative factor $\alpha_{i,\mathcal{Q}}^{\zeta_u}(t)$ for each $\mathbf{v}_{i,\mathcal{Q}}^{\zeta_u}(t)$ similar to Example \ref{example linear neutralization delivery}, each receiver can successfully decode its desired symbols in  $\binom{L-1}{q}(L-q)$-symbol extension, and a per-user DoF of 1 is achieved when $p+q\ge L$.

\begin{remark}\label{remark coded multicast}
Note that coded multicasting is not adopted in the above delivery scheme. Instead, self-interference cancellation is used. This is because if we generate coded messages, each message will be desired by $q+1$ receivers, but will cause interference to the rest $K-q-1$ receivers since this message is not cached at any of them. On the other hand, if we directly transmit each subfile, it is already cached at $q$ out of any $L$ consecutive receivers, hence causes interference to much less receivers. As a result, compared to coded multicasting, this method can create the interference-free transmission for more receivers.
\end{remark}

Now let us consider the case when $p+q<L$. When $p+q<L$, it is easy to verify that each symbol cannot be canceled or neutralized at all unwanted receivers. To tackle this problem, we split the transmission process into $\binom{L}{p+q}$ stages in time such that a distinct set of $p+q$ out of any $L$ consecutive receivers are selected in each stage to be active and need to receive signals from transmitters, while the rest $L-p-q$ receivers do not intend to receive any signal. We aim to create the interference-free transmission for these active receivers in each stage. We represent each stage by a unique combination tuple $(r_1,r_2,\ldots,r_{p+q})$, with $r_i\in [0:L-1]$ and sorted as $0\!\le\! r_1\!<\!\ldots\!<\!r_{p+q}\!\le\! L-1$. In each stage $\Re=(r_1,r_2,\ldots,r_{p+q})$, receivers $\{i:(i\bmod L)\in \Re\}$ are active, while the rest receivers do not need to receive or decode any signal. We split each subfile $W_{i,\mathcal{Q}}^{\zeta_u}$ into $\binom{L-q-1}{p-1}$ equal-sized mini-files, denoted by $\{W_{i,\mathcal{Q},\Re}^{\zeta_u}:\Re\subset[0:L-1],\Re\supseteq(\{i\bmod L\}\cup \mathcal{Q}),|\Re|=p+q\}$. Each mini-file $W_{i,\mathcal{Q},\Re}^{\zeta_u}$ is delivered in stage $\Re$, and each active receiver desires $\binom{p+q-1}{q}(L-q)$ mini-files in each stage. In each stage $\Re$, each mini-file $W_{i,\mathcal{Q},\Re}^{\zeta_u}$ is cached at $q$ out of any $p+q$ adjacent active receivers (since $\mathcal{Q}\subset\Re$), and can be neutralized at the rest undesired active receivers by transmitter cooperation, similar to the previous case when $p\!+\!q\!=\!L$. The transmission strategy when $p\!+\!q\!\ge\! L$ can be directly applied here, and each active receiver can achieve a per-user DoF of 1. Since each receiver is active in $\binom{L-1}{p+q-1}$ stages, the per-user DoF of $\frac{\binom{L-1}{p+q-1}}{\binom{L}{p+q}}\!=\!\frac{p+q}{L}$ is achieved in the entire delivery phase.

Combining two cases when $p+q\ge L$ and when $p+q<L$, the per-user DoF of $\frac{\min\{p+q,L\}}{L}$ is achievable. Since each receiver desires $\binom{L-1}{q}(L-q)$ subfiles, each with $\frac{1}{\binom{L}{q}(L-q)}F$ bits, the achievable NDT at integer-point cache size pair $(\mu_T=\frac{p}{L},\mu_R=\frac{q}{L})$ is
\begin{align}
 \tau(\mu_T=\frac{p}{L},\mu_R=\frac{q}{L})&=\frac{\binom{L-1}{q}(L-q)\cdot\frac{1}{\binom{L}{q}(L-q)}}{\frac{\min\{p+q,L\}}{L}}=\frac{(L-q)}{\min\{p+q,L\}}.\label{eqn artime 2}
\end{align}

\begin{remark}[Computational complexity of the enhanced scheme]\label{remark complexity enhanced}
In the cache placement phase, since we split each file into $(L-q)\binom{L}{q}$ subfiles, the computational complexity is given by $O(NL^{q+1})$. In the content delivery phase, the transmit beamforming vector of each subfile at each time slot is determined by solving a linear equation as in Appendix C, and the overall computational complexity is given by $O(K^4L^{2q+2})$ when $p+q\ge L$ and $O(K^4p^3L^{p+q-2}(p+q)^{2q})$ when $p+q<L$. Similar to the basic scheme, the computational complexity of the receive combining matrix design comes from the inverse operation of square matrices, and is given by $O(KL^{3q+3})$ when $p+q\ge L$ and $O(KL^{p+q+2}(p+q)^{3q})$ when $p+q<L$.
\end{remark}

\subsection{Achievable NDT by Combining the Two Schemes}
Combining the enhanced scheme for large transmitter cache size and the basic scheme for small transmitter cache size in Section \ref{section alignment}, we obtain a lower achievable NDT given as follows.

\begin{theorem}\label{thm achievable NDT combine}
(Achievable NDT-2 for linear network) For the cache-aided $(K+L-1)\times K$ partially connected linear interference network, an achievable NDT-2 is given by
\begin{align}
\tau_{\textrm{A2}}(\mu_T=\frac{p}{L},\mu_R=\frac{q}{L})\triangleq\left\{
\begin{array}{ll}
  \frac{\left(L-1+\frac{L}{q+1}\right)\left(L-q\right)}{L^2}, \quad  &\textrm{ if }p=1, q\in[0:L-1],  \\
  \frac{(L-q)}{\min\{p+q,L\}}, \quad &\textrm{ if } p\in[2:L], q\in[0:L-1].
\end{array}
\right.\label{eqn thm 2}
\end{align}
For general feasible cache size pair $(\mu_T,\mu_R)$ satisfying $L\mu_T+\mu_R\ge1$, the achievable NDT-2 is given by the lower convex envelope of these points and the trivial points $\tau(\mu_T\ge0,\mu_R=1)=0$ using memory sharing.
\end{theorem}

Similar to \eqref{eqn integer tau 1}, we can rewrite \eqref{eqn thm 2} as $\tau=\frac{1-\mu_R}{\min\{\mu_T+\mu_R,1\}}$ when $p\in[2:L],q\in[0:L-1]$, which suggests that the enhanced scheme exploits the receiver local caching gain of $(1-\mu_R)$, and a combined interference neutralization and self-interference cancellation gain of $\frac{1}{\min\{\mu_T+\mu_R,1\}}$. In the latter gain, similar to \cite{bothcache}, $\mu_T$ represents the contribution of interference neutralization gain, and $\mu_R$ represents the contribution of self-interference cancellation gain. In the extreme case when $\mu_T + \mu_R \ge1$, further increasing $\mu_T$ or $\mu_R$ will not increase the gain future. The optimality of NDT-2 in Theorem \ref{thm achievable NDT combine} is stated in the following theorem.

\begin{theorem}\label{thm optimal NDT neutralization}
(Optimality) For the considered network, the achievable NDT-2 is optimal if  $\mu_T+\mu_R\ge1$, with $\tau^*=1-\mu_R$.
\end{theorem}

The proof of Theorem \ref{thm optimal NDT neutralization} is similar to that of Theorem \ref{thm gap}, by comparing the achievable NDT-2 using memory sharing between $(\mu_T=1,\mu_R=0)$ and $(\mu_T=0,\mu_R=1)$ to the lower bound in \eqref{eqn lowerbound 1} in Appendix B, and thus neglected. Compared to Theorem \ref{thm optimal NDT alignment}, Theorem \ref{thm optimal NDT neutralization} shows that the enhanced scheme can achieve the optimal NDT in a larger cache size region than that in the basic scheme, with the expanded region being $1-\mu_T\le\mu_R<\frac{L-1}{L}$. This is because when $1-\mu_T\le\mu_R<\frac{L-1}{L}$, all the interferences can be eliminated at each receiver by using interference neutralization and self-interference cancellation. Thus, all the dimensions of  the received signal space at each receiver are occupied by its desired messages, and we can achieve the optimal NDT of $\tau^*=1-\mu_R$. While our basic scheme for small transmitter cache size is similar to \cite{niesen} in terms of exploiting coded multicasting and interference alignment, the major novelty of our work compared to \cite{niesen} is on the exploitation of self-interference cancellation and interference neutralization in the enhanced scheme for large transmitter cache size, which are ignore in \cite{niesen}.

\begin{remark}[Impact of coding in cache placement]
Note that our proposed caching strategies only involve pure file splitting without any inter-file or intra-file coding. This implies that inter-file or intra-file coding have no benefits on reducing NDT when $\mu_T+\mu_R\ge1$, and can at most reduce NDT by two times in the rest cache size region. This result is related to the problem of multiple unicast sessions in network coding, since our considered receiver demands are multiple unicast. As shown in \cite{networkcoding1,networkcoding2}, network coding can only provide a marginal gain for multiple unicast sessions, while the major benefit comes from multicasting. This explains why coded cache placement cannot reduce NDT by more than a constant factor in our system.
\end{remark}

\begin{figure}[!tbp]
\begin{centering}
\includegraphics[scale=0.32]{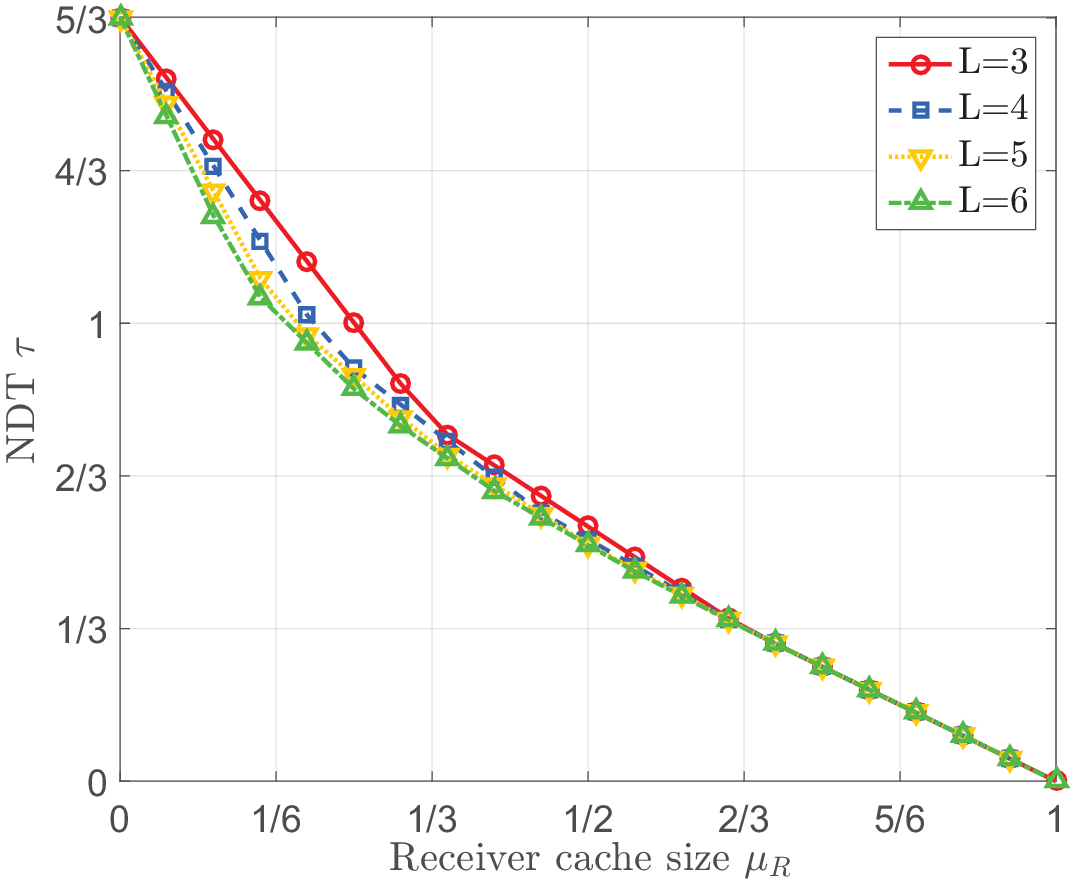}
\caption{Achievable NDT-2 for different receiver connectivity $L$ with $K=10,\mu_T=\frac{1}{3}$.}\label{Fig NDTandL}
\end{centering}
\end{figure}

\begin{remark}[Impact of receiver connectivity]
Note that our achievable NDT-2 in Theorem \ref{thm achievable NDT combine} is only related to the receiver connectivity $L$ but not the receiver number $K$. This is because 1) in the basic scheme, each receiver sees an equivalent $L\times L$ fully connected X-multicast channel by our interference alignment technique; 2) in the enhanced scheme, we create the interference-free transmission for $\min\{p+q,L\}$ out of any $L$ consecutive receivers. To find the impact of receiver connectivity $L$, we plot the achievable NDT-2 under different $L$ with $K=10,\mu_T=\frac{1}{3}$ in Fig. \ref{Fig NDTandL}. It can be seen that the achievable NDTs for all the four different $L$'s are the same and optimal when $\mu_R\ge\frac{2}{3}$ with $\tau^*=1-\mu_R$, as shown in Theorem \ref{thm optimal NDT neutralization}. When $0<\mu_R<\frac{2}{3}$, Fig. \ref{Fig NDTandL} shows that the achievable NDT slightly decreases when $L$ increases. This implies that when $L$ increases, the positive impact of having more transmitter cooperation or coordination slightly prevails the negative impact of more interferences received at each receiver. In the extreme case when $\mu_R=0$, the achievable NDT for all the four different $L$'s are the same, i.e., $\tau=\frac{5}{3}$, which implies that the gain of having more transmitter cooperation/coordination is completely offset by the loss due to more interferences.
\end{remark}

\subsection{Comparison with Other Works}\label{sec comparison}
\begin{figure}[!tbp]
\begin{centering}
\includegraphics[scale=0.33]{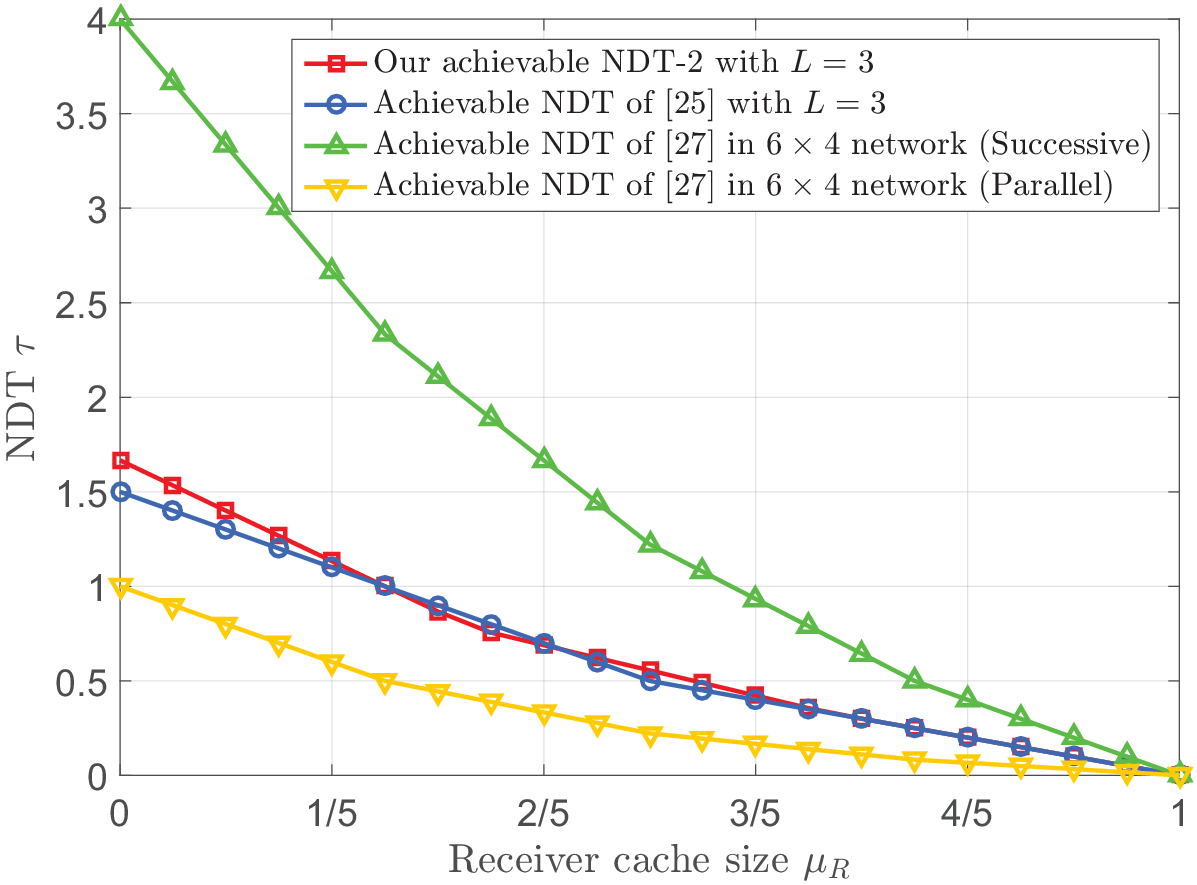}
\caption{Achievable NDTs of our scheme and \cite{Wynercaching} with $L=3$ and \cite{gunduz_random} in the $6\times4$ linear network when $\mu_T=\frac{1}{3}$.}\label{Fig comparison}
\end{centering}
\vspace{-10pt}
\end{figure}

In this subsection, we compare our achievable NDT-2 in Theorem \ref{thm achievable NDT combine} with \cite{Wynercaching,gunduz_random}. The authors in \cite{Wynercaching} considered the Wyner's circular model with $L=2$ or $3$, and obtained the per-receiver rate-memory tradeoff. The authors in \cite{gunduz_random} considered a random network topology with caches at all transmitters and receivers, and studied both successive and parallel transmissions.

In Fig. \ref{Fig comparison}, we plot the achievable NDTs of our scheme and \cite{Wynercaching} with $L=3$, since both NDTs are only related to $L$ but not receiver number $K$, and the achievable NDT of \cite{gunduz_random} in the $6\times4$ linear network with $L=3$, when $\mu_T=\frac{1}{3}$. Note that each transmitter in \cite{Wynercaching} can download all the desired files of its connected receivers from the database after the receiver request is revealed. Thus, it is natural that our achievable NDT is worse, though only slightly, than that in \cite{Wynercaching} when $\mu_R$ is small. However, our achievable NDT is still close to it in the entire cache size region. As for \cite{gunduz_random}, the achievable NDT for parallel transmission is the best among all NDTs, because the messages are transmitted through dedicated links from transmitters to the connected receivers without interference. However, by using successive transmission, their achievable NDT is much worse since transmitters take turns to transmit messages to avoid interferences. The above results demonstrate that our proposed caching and delivery schemes are effective.

\section{Extension to Circular Networks and Heterogeneous Linear Networks}\label{sec circular}
\subsection{Circular Networks}
As mentioned in Section II-A, our partially connected linear interference network has strong connection with the circular interference network. Recall that if we merge transmitter pair $(j, K+j)$, for $j \in [0:L-2]$, in the $(K+L-1) \times K$ linear network, we will arrive at a $K \times K$ circular network. To make the two network models completely identical, the cached contents on the merged node should remain the same as before merging. By the proposed caching strategies, the cached contents in the transmitter pair $(j, K+j)$ of the linear network are identical when $L$ is a divisor of $K$.  Thus, under this condition, the achievable NDT-2 in Theorem \ref{thm achievable NDT combine} is directly applicable to the $K \!\times\! K$ circular network.

In the special case when $L=K$, the circular network becomes the fully connected network. Then, the achievable per-user DoFs of $\frac{L}{L-1+\frac{L}{q+1}}$ in Lemma \ref{lemma alignment dof} and $\frac{\min\{p+q,L\}}{L}$ in the enhanced scheme are consistent with \cite[Theorem 2]{niesen} and \cite[Theorem 1]{bothcache}, respectively.

\subsection{Linear Networks with Heterogeneous Receiver Connectivity}\label{sec hete}
Recall that each receiver $i$, for $i\in[0:K-1]$, is connected to $L_i$ consecutive transmitters $[i:i+L_i-1]$ in the heterogeneous linear network. Due to the heterogeneous receiver connectivity, our proposed schemes are not directly applicable to this network. In this subsection, we aim to adjust our basic scheme for this network by focusing on cache size pairs $(\mu_T=\frac{1}{L_m},\mu_R=\frac{q}{K})$ with $L_m\!\triangleq\!\min\limits_{i\in[0:K-1]}\!L_i$ and $q\in[0:K-1]$.

\subsubsection{Cache placement}
Split each file $W_n$ into $L_m\binom{K}{q}$ equal-sized subfiles, denoted by $\{W_{n,\mathcal{Q}}^\zeta\!:\!\mathcal{Q}\!\subset\![0\!:\!K-1],|\mathcal{Q}|\!=\!q,\zeta\!\in\![0\!:\!L_m-1]\}$. Each subfile $W_{n,\mathcal{Q}}^\zeta$ is cached at receiver set $\{i\!:\!i\!\in\! \mathcal{Q}\}$ and transmitter set $\{j\!:\!(j\bmod L_m)\!=\! \zeta\}$. Note that this cache placement is cyclic at transmitters but not at receivers  due to the heterogeneous receiver connectivity.

\subsubsection{Content delivery}
We still assume that each receiver $i$ desires file $W_i$. Since each subfile desired by one receiver is already cached at $q$ receivers, coded multicasting can be adopted similar to Section \ref{sec alignment delivery}. Specifically, each transmitter $j$ generates the following coded messages
\begin{align}
  &\left\{W_{\textrm{Tx}_j}^\mathcal{R}=\bigoplus_{i\in\mathcal{R}}W_{i,\mathcal{R}\setminus\{i\}}^{\hat{j}}: \mathcal{R}\in[0:K-1], |\mathcal{R}|=q+1, \mathcal{R}\cap\mathcal{R}^h_j\neq \emptyset\right\},\label{eqn hete coded message}
\end{align}
where $\hat{j}=j\bmod L_m$ and $\mathcal{R}^h_j$ denotes the set of receivers connected to transmitter $j$. Each message $W_{\textrm{Tx}_j}^\mathcal{R}$ in \eqref{eqn hete coded message} is needed by receivers in $\mathcal{R}\cap\mathcal{R}^h_j$.  Note that for each receiver $i$ and each receiver set $\mathcal{R}\ni i$, if $L_i>L_m$, then transmitters connected to receiver $i$ and with indices congruent modulo $L_m$ transmit the same message intended for set $\mathcal{R}$, i.e., $W_{\textrm{Tx}_{j_1}}^\mathcal{R}=W_{\textrm{Tx}_{j_2}}^\mathcal{R}$ for $j_1,j_2\in[i:i+L_i-1]$ and $j_1\equiv j_2\; (\bmod\; L_m)$.

The transmission method of these messages is very similar to Appendix A, where the only difference lies on the message partition. Here, all the messages are divided into $\binom{K}{q+1}$ message sets according to their intended receiver set $\mathcal{R}$, i.e., message set $\mathcal{W}_\mathcal{R}$, for $\mathcal{R}\in[0:K-1],|\mathcal{R}|=q+1$, contains messages $\{W_{\textrm{Tx}_j}^{\mathcal{R}}\mid j\in [0:K_T-1],\mathcal{R}\cap\mathcal{R}^h_j\neq\emptyset\}$ with $K_T$ being the number of transmitters in the network. Messages in  the same set are  aligned in a same subspace at undesired receivers. Due to page limit, the detailed transmission method is neglected here. For each receiver $i$, it can successfully decode its $L_m$ desired messages in each message set $\mathcal{W}_{\mathcal{R}}$ with $\mathcal{R}\ni i$ \footnote{As mentioned before, transmitters connected to receiver $i$ and with indices congruent modulo $L_m$ transmit the same message in $\mathcal{W}_{\mathcal{R}}$ with $\mathcal{R}\ni i$. Hence, each receiver only has $L_m$ desired and independent messages in this set $\mathcal{W}_{\mathcal{R}}$.}, while its undesired messages in each set $\mathcal{W}_{\mathcal{R}}$ with $\mathcal{R}\not\ni i$ are aligned together. Hence, a per-user DoF of $d=\frac{L_m\binom{K-1}{q}}{L_m\binom{K-1}{q}+\binom{K-1}{q+1}}=\frac{L_m}{L_m-1+\frac{K}{q+1}}$ is achieved. Since each receiver desires $L_m\binom{K-1}{q}$ subfiles, each with $\frac{F}{L_m\binom{K}{q}}$ bits, the achievable NDT is given by $\tau=\frac{L_m\binom{K-1}{q}}{L_m\binom{K}{q}}\cdot \frac{1}{d}=\frac{(L_m-1+\frac{K}{q+1})(K-q)}{L_mK}$.

\subsubsection{Achievable NDT}
The achievable NDT is formally stated in the following proposition.
\begin{proposition}\label{thm hete achievable NDT}
(Achievable NDT) For the cache-aided heterogeneous linear network as defined above, an achievable NDT is given by
\begin{align}
\tau_{\textrm{A}}^h(\mu_T\ge\frac{1}{L_m},\mu_R=\frac{q}{K})\triangleq &\frac{\left(L_m-1+\frac{K}{q+1}\right)\left(K-q\right)}{L_mK}, \quad  \textrm{ if } q\in[0:K-1],\label{eqn thm hete 1}
\end{align}
with $L_m\!=\!\min_{i\in[0:K-1]}\!L_i$. For general feasible cache size pair $(\mu_T,\mu_R)$ satisfying $L_m\mu_T+\mu_R\ge1$, the achievable NDT is given by the lower convex envelope of these points  and the trivial points $\tau(\mu_T\ge0,\mu_R=1)=0$ using memory sharing.
\end{proposition}
\begin{proposition}\label{thm hete gap}
(Multiplicative Gap) For the heterogeneous linear network, the multiplicative gap between the achievable NDT in Proposition \ref{thm hete achievable NDT} and the optimal NDT is less than $\frac{L_m-1+K}{L_m}$.
\end{proposition}
The proof of Proposition \ref{thm hete gap} is the same as that of Theorem \ref{thm gap}, and thus neglected. By adjusting our basic scheme for this network model, the caching strategy, the transmit beamforming vector design, and the receive combining matrix design have a computational complexity of $O(NK^{q})$, $O(K_T^2K^{2q+2}n^{2K_T(K-q-1)})$, and $O(K^{3q+1}n^{3K_T(K-q-1)})$, respectively.

\section{Conclusion and Future Direction} \label{section conclusion}
In this paper, we studied caching in  partially connected linear  networks. We first proposed a basic scheme where each of every $L$ consecutive transmitters caches a distinct part of each file in the cache placement phase. In the delivery phase, by introducing virtual receivers, we exploited both coded multicasting gain and interference alignment gain. The achievable NDT is within a multiplicative gap of 2 to the optimum, and optimal when $\mu_R\ge\max\{1-\mu_T,\frac{L-1}{L}\}$. To fully utilize the transmitter cache and hence obtain a lower NDT, we then proposed an enhanced scheme, where each subfile is cached repeatedly at $p$ consecutive transmitters. By utilizing the overlapped transmitter cache, we can exploit both self-interference cancellation gain and interference neutralization gain in the delivery phase. By combining these two schemes, the achievable NDT is optimal when $\mu_T+\mu_R\ge1$. Finally, we extended our schemes to the circular network when $L$ divides $K$, and also extended our basic scheme to the linear network with heterogeneous receiver connectivity.

Among open problems, since our proposed schemes suffer from large computational complexity, how to design proper schemes with low complexity is an important research direction. In addition, given that this paper assumes a central controller to collect the global CSI, it is still of great importance to study achievable schemes when there is no central controller and only local CSI is available at transmitters and receivers.

\section*{Appendix A: Proof of Lemma \ref{lemma alignment dof}}\label{appendix alignment proof}

We divide all the messages into $\binom{L}{q+1}$ message sets. Each message set is represented by a unique combination tuple $(r_0,r_1,\ldots,r_{q})$, with $r_i\in [0:L-1]$ and sorted as $0\le r_0<\ldots<r_{q}\le L-1$. Messages in an arbitrary message set $\mathcal{\Re}=(r_0,r_1,\ldots,r_{q})$ are given by
\begin{align}
\mathcal{W}_{\mathcal{\Re}}=&\left\{W_{\textrm{Tx}_j}^{\mathcal{R}}\mid j\in [0:K+L-2],\mathcal{R}\cap[0:K-1]\neq\emptyset,\mathcal{R}\subseteq \mathcal{R}_j^e, \mathcal{R}\bmod L=\mathcal{\Re}\right\}.\label{eqn:message group}
\end{align}
From \eqref{eqn:message group}, it can be seen that each transmitter has at most one message in each set. Messages in the same set will be aligned in a same subspace at undesired receivers. This interference management idea can also be found similarly in \cite{Xchannel,niesen}, but differs in message partition scheme.

We use a $T_n=L\binom{L-1}{q}n^{\rho}+\binom{L-1}{q+1}(n+1)^{\rho}$-symbol extension, with $n\in\mathbb{Z}^+,\rho\triangleq(K+L-1)(L-q-1)$, and the channel between transmitter $j$ and receiver $i$ ($j\in \mathcal{T}_i$) becomes a $T_n\times T_n$ diagonal matrix $\mathbf{H}_{i,j}$ whose diagonal entries $h_{i,j}(t)$ ($1\le t\le T_n$) are i.i.d. following some continuous distribution. We encode each message $W_{\textrm{Tx}_j}^{\mathcal{R}}$ into a column vector of $n^{\rho}$ symbols $\mathbf{x}_{j}^{\mathcal{R}}=[x_{j,m}^{\mathcal{R}}]_{1\le m\le n^{\rho}}$, and  each symbol $x_{j,m}^{\mathcal{R}}$ is beamformed along a $T_n\times1$ column vector $\mathbf{v}_{j,m}^{\mathcal{R}}=\left[v_{j,m}^{\mathcal{R}}(t)\right]_{1\le t\le T_n}$. Then, the codeword of message $W_{\textrm{Tx}_j}^{\mathcal{R}}$ is
$\mathbf{V}_{j}^{\mathcal{R}}\mathbf{x}_{j}^{\mathcal{R}}\triangleq\sum_{m}\mathbf{v}_{j,m}^{\mathcal{R}}x_{j,m}^{\mathcal{R}}$. The received signal at an arbitrary actual receiver $i$, for $i\in[0:K-1]$, is given by (neglecting the noise)
\begin{align}
\mathbf{y}_i&=\sum_{j\in \mathcal{T}_i}\mathbf{H}_{i,j}\sum_{\substack{\mathcal{R}:\mathcal{R}\cap[0:K-1]\neq\emptyset,\\\mathcal{R}\subseteq\mathcal{R}_j^e,|\mathcal{R}|=q+1}}\mathbf{V}_{j}^{\mathcal{R}}\mathbf{x}_{j}^{\mathcal{R}}=\sum_{j\in \mathcal{T}_i}\mathbf{H}_{i,j}\sum_{\substack{\mathcal{R}:\mathcal{R}\cap[0:K-1]\neq\emptyset,\\\mathcal{R}\subseteq\mathcal{R}_j^e,|\mathcal{R}|=q+1}}\sum_{m=1}^{n^{\rho}}\mathbf{v}_{j,m}^{\mathcal{R}}x_{j,m}^{\mathcal{R}}.\label{eqn received orginal signal}
\end{align}

In \eqref{eqn received orginal signal}, symbols $\{x_{j,m}^{\mathcal{R}}\}$ satisfying $i\notin \mathcal{R}$ are undesired by receiver $i$, and should be aligned. Specifically, symbols in the same set \eqref{eqn:message group} are aligned together. Consider channel matrix set
\begin{align}
\mathcal{H}_{\mathcal{\Re}}=&\left\{\mathbf{H}_{i,j}\mid j\in [0:K+L-2], \mathcal{R}\subseteq\mathcal{R}_j^e,\mathcal{R}\bmod L=\mathcal{\Re},i\in\mathcal{R}_j^e\backslash\mathcal{R} \right\}\label{eqn:interference matrix}
\end{align}
for each message set $\mathcal{\Re}$. There are $\rho$ matrices in $\mathcal{H}_{\mathcal{\Re}}$. Note that the channel matrices observed in virtual receivers from their connected transmitters are also assumed to be diagonal and independent with each other and the original channel matrix. The diagonal entries $h_{i,j}(t)$ ($1\!\le\! t\!\le\! T_n$) of channel matrix between each virtual receiver and its connected transmitter are i.i.d. following some continuous distribution. Then, for each message $W_{\textrm{Tx}_j}^{\mathcal{R}}\!\in\!\mathcal{W}_{\mathcal{\Re}}$, the beamforming vector $\mathbf{v}_{j,m}^{\mathcal{R}}$ of each $x_{j,m}^{\mathcal{R}}$ is given by a unique vector in vector set
\begin{equation}
\mathcal{V}_{\mathcal{\Re}}(n)=\bigg\{\prod_{H_{i,j}\in\mathcal{H}_{\mathcal{\Re}}}\left(\mathbf{H}_{i,j}\right)^{\alpha_{i,j}}\cdot \mathbf{b}_{\mathcal{\Re}}\mid 1\le\alpha_{i,j}\le n\bigg\},\label{eqn appendix a matrix set}
\end{equation}
where $\mathbf{b}_{\mathcal{\Re}}$ is a $T_n\times 1$ column vector $[b_{\mathcal{\Re}}(t)]_{1\le t\le T_n}$ whose entries are i.i.d. following some continuous distribution, and independent with other $\mathbf{b}_{\mathcal{\Re}}$ for different $\mathcal{\Re}$. Then, we can rewrite \eqref{eqn received orginal signal} as
\begin{equation}
\mathbf{y}_i=\sum_{j\in\mathcal{T}_i}\mathbf{H}_{i,j}\sum_{\mathcal{\Re}}\sum_{m=1}^{n^{\rho}}\prod_{H_{i,j}\in\mathcal{H}_{\mathcal{\Re}}}\left(\mathbf{H}_{i,j}\right)^{\alpha_{i,j}(m)}\mathbf{b}_{\mathcal{\Re}}x_{j,m}^{\mathcal{R}},\label{eqn:received signal}
\end{equation}
where $\alpha_{i,j}(m)$ denotes the power of $\mathbf{H}_{i,j}$ for symbol $x_{j,m}^{\mathcal{R}}$. For undesired message $W_{\textrm{Tx}_j}^{\mathcal{R}}\in\mathcal{W}_{\mathcal{\Re}}$ where $i\notin \mathcal{R}$ in \eqref{eqn:received signal} (thus $(i\bmod L)\notin \mathcal{\Re}$), it can be seen that $\mathbf{H}_{i,j}\in\mathcal{H}_{\mathcal{\Re}}$. Hence, the received beamforming vector $\mathbf{H}_{i,j}\prod\left(\mathbf{H}_{i,j}\right)^{\alpha_{i,j}}\mathbf{b}_{\mathcal{\Re}}$ of $x_{j,m}^{\mathcal{R}}$ lies in vector set $\mathcal{V}_{\mathcal{\Re}}(n+1)$. Interferences are thus aligned, i.e., the received beamforming vectors of undesired messages $\{W_{\textrm{Tx}_j}^{\mathcal{R}}:W_{\textrm{Tx}_j}^{\mathcal{R}}\in\mathcal{W}_{\mathcal{\Re}},(i\bmod L)\notin \mathcal{\Re},j\in\mathcal{T}_i\}$ all lie in $\mathcal{V}_{\mathcal{\Re}}(n+1)$.

The received beaforming vectors of desired messages of receiver $i$ are given by the column vectors of the following $T_n\times L\binom{L-1}{q}n^{\rho}$ matrix:
\begin{align}
\mathbf{D}_i=\left[\mathbf{H}_{i,j}\mathbf{V}_{j}^{\mathcal{R}}\right]_{j \in\mathcal{T}_i, \textrm{ and } \mathcal{R} \textrm{ satisfies } i\in\mathcal{R}}.
\end{align}
The received beamforming vectors of undesired messages of receiver $i$ lie in the linear space spanned by the vectors in $\mathcal{V}_{\mathcal{\Re}}(n+1)$ for $\mathcal{\Re}$ satisfying $(i\bmod L)\notin \mathcal{\Re}$. This is equivalent to the linear space spanned by the column vectors of the following $T_n\times\binom{L-1}{q+1}(n+1)^{\rho}$ matrix:
\begin{align}
\mathbf{I}_i=\left[\mathbf{u}\right]_{u\in\mathcal{V}_{\mathcal{\Re}}(n+1), \textrm{ and } \mathcal{\Re} \textrm{ satisfies } (i\bmod L)\notin \mathcal{\Re}}.
\end{align}
We need to assure that the $T_n\times T_n$ matrix $\mathbf{A}_i\!\triangleq\!\left[\mathbf{D}_i\;\;\mathbf{I}_i\right]$ is full rank almost surely. By using \cite[Lemma 1]{Xchannel}, we only need to prove the entries in the same row of $\mathbf{A}_i$ are different monomials. Consider an arbitrary $t$-th row of $\mathbf{A}_i$. It is obvious that entries of messages in different message sets $\mathcal{\Re}$ differ in $b_{\mathcal{\Re}}(t)$. Then consider an arbitrary set $\mathcal{\Re}$. We first consider the set $\mathcal{\Re}$ such that $(i\bmod L)\in\mathcal{\Re}$. In this set, each transmitter $j\in\mathcal{T}_i$ has a message $W_{\textrm{Tx}_j}^{\mathcal{R}}\in\mathcal{W}_{\mathcal{\Re}}$ desired by receiver $i$. Since $\mathbf{H}_{i,j}\notin\mathcal{H}_{\mathcal{\Re}}$, entries of each $W_{\textrm{Tx}_j}^{\mathcal{R}}$ in this set have a unique term $h_{i,j}(t)$. Finally, consider an arbitrary $W_{\textrm{Tx}_j}^{\mathcal{R}}$. Given the design in \eqref{eqn appendix a matrix set}, its corresponding entries are different monomials. Similar arguments can be applied to message sets $\mathcal{\Re}$ satisfying $(i\bmod L)\notin\mathcal{\Re}$.  Thus we proved that the entries in the same row of  $\mathbf{A}_i$ are different monomials. By \cite[Lemma 1]{Xchannel}, we assure that $\mathbf{A}_i$ is full rank almost surely. Then, the receive combining matrix for receiver $i$ is given by $\mathbf{A}_i^{-1}$. Thus, receiver $i$ can successfully decode its desired symbols by computing $\mathbf{A}_i^{-1}\mathbf{y}_i$, so as the other receivers.

Since each receiver decodes its $L\binom{L-1}{q}$ desired messages, each encoded into $n^{\rho}$ symbols, in $T_n$-symbol extension, we can achieve a per-user DoF of $\frac{L\binom{L-1}{q}n^{\rho}}{L\binom{L-1}{q}n^{\rho}+\binom{L-1}{q+1}(n+1)^{\rho}}$. Letting $n\rightarrow \infty$, a per-user DoF of $\frac{L\binom{L-1}{q}}{L\binom{L-1}{q}+\binom{L-1}{q+1}}=\frac{L}{L-1+\frac{L}{q+1}}$ is achieved. Lemma \ref{lemma alignment dof} is proved.

\section*{Appendix B: Multiplicative Gap of NDT (Proof of Theorem \ref{thm gap})}
First, let us present a lower bound of the optimal NDT in the following lemma.
\begin{lemma}\label{thm converse NDT linear}
For the $(K+L-1)\times K$ linear network, the optimal NDT is lower bounded by
\begin{align}
\tau^*\ge\tau_{\textrm{lb}}\triangleq1-\mu_R.\label{eqn lowerbound 1}
\end{align}
\end{lemma}
The proof is based on a cut-set argument, similar to \cite{fundamentallimits}, and thus ignored. For any feasible cache size pair $(\mu_T,\mu_R)$, we can use memory sharing between $(\mu_T=0,\mu_R=1)$ and $(\mu_T=\frac{1}{L},\mu_R=0)$ to obtain an achievable NDT. Specifically, we split each file into two subfiles. One has $\mu_RF$ bits, and is cached at all receivers but none of the transmitters, as at $(\mu_T=0,\mu_R=1)$, yielding $\tau_1=0$. The other has $(1-\mu_R)F$ bits, and is cached and delivered by the basic scheme at $(\mu_T=\frac{1}{L},\mu_R=0)$, yielding $\tau_2=\frac{2L-1}{L}$. By memory sharing, the corresponding achievable NDT is $\tau=\frac{(2L-1)(1-\mu_R)}{L}$. Comparing to \eqref{eqn lowerbound 1}, we have  $\frac{\tau_{\textrm{A1}}}{\tau^*}\le\frac{2L-1}{L}<2$. Theorem \ref{thm gap} is proved.

\section*{Appendix C: Design of Beamforming Vectors when $p+q\ge L$}
We consider vector $\mathbf{v}_{i_0,\mathcal{Q}}^{\zeta_u}(t)$ of an arbitrary symbol $x_{i_0,\mathcal{Q}}^{\zeta_u}$ at an arbitrary time slot $t$ to illustrate the beamforming design. Note that $x_{i_0,\mathcal{Q}}^{\zeta_u}$ is desired by receiver $i_0$, and available at transmitters $\{j:(j\bmod L)\in[\zeta_u+L-p:\zeta_u+L-1]\bmod L\}$. Define $\zeta_u^{i_0}\triangleq \lfloor\frac{i_0-\zeta_u}{L}\rfloor L+\zeta_u$ as the maximum number congruent with $\zeta_u$ modulo $L$ which is no larger than $i_0$. Note that $W_{i_0,\mathcal{Q}}^{\zeta_u}$ is cached at transmitters $[\zeta_u^{i_0}+L-p:\zeta_u^{i_0}+L-1]$ with $\zeta_u^{i_0}+L-1\in[i_0:i_0+L-1]$. We let these $p$ transmitters deliver $x_{i_0,\mathcal{Q}}^{\zeta_u}$ to receiver $i_0$, and neutralize it at the unwanted receivers $[\zeta_u^{i_0}:\zeta_u^{i_0}+L-1]\backslash\{i_0\}$.

The channel matrix between transmitters $[\zeta_u^{i_0}+L-p:\zeta_u^{i_0}+L-1]$ and receivers $[\zeta_u^{i_0}:\zeta_u^{i_0}+L-1]$ at each time slot $t$ is given by
\begin{equation}
\mathbf{H}_{[\zeta_u^{i_0}+L-p:\zeta_u^{i_0}+L-1]}^{[\zeta_u^{i_0}:\zeta_u^{i_0}+L-1]}\!(t)\!=\!\!\left[\!\!
  \begin{array}{cccc}
    h_{\zeta_u^{i_0},\zeta_u^{i_0}+L-p}\!(t) & h_{\zeta_u^{i_0},\zeta_u^{i_0}+L-p+1}\!(t) & \cdots & h_{\zeta_u^{i_0},\zeta_u^{i_0}+L-1}\!(t) \\
    \cdots & \cdots & \cdots & \cdots \\
    h_{\zeta_u^{i_0}+L-p,\zeta_u^{i_0}+L-p}\!(t) & h_{\zeta_u^{i_0}+L-p,\zeta_u^{i_0}+L-p+1}\!(t) & \cdots & h_{\zeta_u^{i_0}+L-p,\zeta_u^{i_0}+L-1}\!(t) \\
    0 & h_{\zeta_u^{i_0}+L-p+1,\zeta_u^{i_0}+L-p+1}\!(t) & \cdots & h_{\zeta_u^{i_0}+L-p+1,\zeta_u^{i_0}+L-1}\!(t) \\
    0 & 0 & \cdots & \cdots \\
    0 & 0 & 0 & h_{\zeta_u^{i_0}+L-1,\zeta_u^{i_0}+L-1}\!(t) \\
  \end{array}
\!\!\right]\!\!.\label{eqn channel matrix}
\end{equation}
It can be seen that the channel matrix of each symbol has the same form as in \eqref{eqn channel matrix} by using the modified CC strategy, which largely reduces the complexity of precoding coefficient design in our scheme. Among these $L$ receivers, $q$ receivers already cache subfile $W_{i_0,\mathcal{Q}}^{\zeta_u}$, and can cancel $x_{i_0,\mathcal{Q}}^{\zeta_u}$ using its local cache. Thus, we only need to design the precoding coefficients at transmitters $[\zeta_u^{i_0}+L-p:\zeta_u^{i_0}+L-1]$ to neutralize $x_{i_0,\mathcal{Q}}^{\zeta_u}$ at the rest $L-q-1$ unwanted receivers, and ensure that the received factor of $x_{i_0,\mathcal{Q}}^{\zeta_u}$ is non-zero at the desired receiver $i_0$. Given that these $L-q$ receivers correspond to random $L-q$ rows in  $\mathbf{H}_{[\zeta_u^{i_0}+L-p:\zeta_u^{i_0}+L-1]}^{[\zeta_u^{i_0}:\zeta_u^{i_0}+L-1]}(t)$, it suffices to prove that $\mathbf{H}_{[\zeta_u^{i_0}+L-p:\zeta_u^{i_0}+L-1]}^{[\zeta_u^{i_0}:\zeta_u^{i_0}+L-1]}(t)$ is full row rank with probability one for arbitrary $L-q$ rows. This is true since each channel coefficient is i.i.d. following some continuous distribution, and the diagonal entries of the matrix formed by arbitrary $L-q$ rows and the last $L-q$ columns are non-zero (which implies that the determinant of this $(L\!-\!q)\!\times \!(L\!-\!q)$ matrix is non-zero with probability one). Therefore, there always exists a vector $\big[\tilde{v}_{i_0,\mathcal{Q}}^{\zeta_u,\zeta_u^{i_0}+L-p}(t),\ldots,\tilde{v}_{i_0,\mathcal{Q}}^{\zeta_u,\zeta_u^{i_0}+L-1}(t)\big]^\top$ satisfying
\begin{equation}
  \left\{
  \begin{array}{ll}
  \sum_{j=1}^p h_{i_0,\zeta_u^{i_0}+L-j}(t)\tilde{v}_{i_0,\mathcal{Q}}^{\zeta_u,\zeta_u^{i_0}+L-j}(t)\neq0, \ \textrm{at receiver $i_0$ ,}\\
    \sum_{j=1}^p h_{i,\zeta_u^{i_0}+L-j}(t)\tilde{v}_{i_0,\mathcal{Q}}^{\zeta_u,\zeta_u^{i_0}+L-j}(t)=0, \ \textrm{at $L-q$ unwanted and uncached receivers.}
  \end{array}
  \right.\label{eqn linear neutralization 2}
\end{equation}
The precoding coefficient at transmitter $j$, for $j\!\in\![\zeta_u^{i_0}\!+\!L\!-\!p\!:\!\zeta_u^{i_0}\!+\!L\!-\!1]$, is given by $v_{i_0,\mathcal{Q}}^{\zeta_u,j}(t)=\alpha_{i_0,\mathcal{Q}}^{\zeta_u}(t)\tilde{v}_{i_0,\mathcal{Q}}^{\zeta_u,j}(t)$, where $\alpha_{i_0,\mathcal{Q}}^{\zeta_u}$ is i.i.d. following some continuous distribution, for arbitrary $i_0,\mathcal{Q},\zeta_u$.

Note that $x_{i_0,\mathcal{Q}}^{\zeta_u}$ still causes interference to the unwanted receivers $[\zeta_{u}^{i_0}+1-p:\zeta_u^{i_0}-1]\subseteq[\zeta_u^{i_0}-L:\zeta_u^{i_0}-1]$. Then, transmitters $[\zeta_u^{i_0}-p:\zeta_u^{i_0}-1]$ can similarly cooperate to neutralize $x_{i_0,\mathcal{Q}}^{\zeta_u}$ at the $L-q$ uncached receivers in $[\zeta_u^{i_0}-L:\zeta_u^{i_0}-1]$. Iteratively, $x_{i_0,\mathcal{Q}}^{\zeta_u}$ can be neutralized or canceled at all unwanted receivers in $[0:\zeta_{u}^{i_0}+L-1]\backslash \{i_0\}$. On the other hand, by simply letting $v_{i_0,\mathcal{Q}}^{\zeta_u,j}(t)=0$ for transmitters $[\zeta_{u}^{i_0}+L:K+L-1]$, $x_{i_0,\mathcal{Q}}^{\zeta_u}$ will not cause interference to receivers  $[\zeta_{u}^{i_0}+L:K-1]$. Therefore, symbol $x_{i_0,\mathcal{Q}}^{\zeta_u}$ is only received by its desired receiver $i_0$, so as the other symbols. Thus, an interference-free transmission is created for each receiver.

Since each desired symbol $x_{i_0,\mathcal{Q}}^{\zeta_u}$ of each receiver $i_0$ has a unique factor $\alpha_{i_0,\mathcal{Q}}^{\zeta_u}(t)$ for $t\!\in\![\binom{L\!-\!1}{q}\!(L\!-\!q)]$, it is easy to see that the $\binom{L\!-\!1}{q}\!(L\!-\!q)\!\times\! 1$ received  beamforming vector of each symbol by combining all time slots is linearly independent with each other. Thus, each receiver can  decode its $\binom{L\!-\!1}{q}\!(L\!-\!q)$ desired symbols in $\binom{L\!-\!1}{q}\!(L\!-\!q)$-symbol extension to achieve a per-user DoF of 1.

\bibliographystyle{IEEEtran}
\bibliography{IEEEabrv,journal}

\end{document}